\documentstyle[pra,aps,epsfig,multicol]{revtex}

\begin{document}
\title{First-principles calculations of the structural, 
electronic, vibrational and magnetic properties of 
${\rm C}_{60}$ and ${\rm C}_{48}{\rm N}_{12}$: a comparative 
study}
\author{ Rui-Hua Xie and Garnett W. Bryant} 
\address{Atomic Physics Division, National Institute of 
Standards and Technology, Gaithersburg, MD 20899-8423}
\author{Lasse Jensen}
\address{Theoretical Chemistry, Material Science Centre,
Rijksuniversiteit Groningen, Nijenborgh 4,\\  9747 AG Groningen,  The
Netherlands}
\author{Jijun Zhao}
\address{Department of Physics and Astronomy, University of North
Carolina at Chapel Hill, Chapel Hill, NC 27599}
\author{Vedene H. Smith, Jr.}
\address{Department of Chemistry, Queen's University, Kingston,
ON K7L 3N6, Canada}
\date{\today}
\maketitle
\begin{abstract}

We perform first-principles calculations of the structural, electronic, 
vibrational and magnetic properties of the ${\rm C}_{48}{\rm N}_{12}$ azafullerene and 
${\rm C}_{60}$.  Full geometrical optimization shows that 
${\rm C}_{48}{\rm N}_{12}$ is characterized by several 
distinguishing features: only one nitrogen atom per pentagon, two nitrogen 
atoms preferentially sitting  in one hexagon, ${\rm S}_{6}$ 
symmetry, 6 unique nitrogen-carbon  and 9 unique carbon-carbon 
bond lengths. The Mulliken charge analysis indicates that the doped nitrogen atoms 
in ${\rm C}_{48}{\rm N}_{12}$  exist as electron acceptors and three-fourths 
of the carbon atoms as electron donors. Electronic structure calculations of 
${\rm C}_{48}{\rm N}_{12}$ show that the highest occupied molecular 
orbital (HOMO) is a doubly degenerate level 
of $a_{g}$ symmetry and the lowest unoccupied molecular orbital  (LUMO) is 
a nondegenerate level of  $a_{u}$ symmetry. The calculated binding energy per atom 
and HOMO-LUMO energy gap of ${\rm C}_{48}{\rm N}_{12}$ are about  1 eV smaller 
than those of ${\rm C}_{60}$. For both ${\rm C}_{48}{\rm N}_{12}$ and 
${\rm C}_{60}$, the total energies calculated with STO-3G, 3-21G and 
6-31G basis sets differ from the 6-31G* basis set results by about 
1.5\%, 0.6\% and 0.05\%, respectively. Because of electron correlations, 
the HOMO-LUMO gap decreases about 5 eV and the binding energy per atom 
increases about 2 eV. Our vibrational frequency analysis predicts that 
${\rm C}_{48}{\rm N}_{12}$ has in total 116 vibrational modes: 58 modes 
are infrared-active (29 doubly-degenerate and 29 non-degenerate modes) 
and  58 modes are Raman-active (29 doubly-degenerate unpolarized and 29 non-degenerate 
polarized).   It is found that ${\rm C}_{48}{\rm N}_{12}$ exhibits eight $^{13}{\rm C}$ 
and two $^{15}{\rm N}$ NMR (nuclear magnetic  resonance) spectral signals.  In 
comparison to isolated carbon or nitrogen atoms, an enhancement in the 
dipole polarizability is found due to the delocalized $\pi$ 
electrons in ${\rm C}_{48}{\rm N}_{12}$ and ${\rm C}_{60}$. 
The average second-order hyperpolarizability of ${\rm C}_{48}{\rm N}_{12}$ is about 
55\% larger than that of ${\rm C}_{60}$.  In addition, 
the effects of basis sets are discussed in detail, and the different methods  for 
calculating nuclear magnetic shielding tensors  are compared. 
Our detailed study of ${\rm C}_{60}$ reveals the importance of 
electron correlations and the choice of basis sets in the 
first-principles calculations. Our best-calculated results for 
${\rm C}_{60}$ with the B3LYP hybrid density 
functional theory are in excellent agreement with experiment and 
demonstrate the desired efficiency and accuracy needed  for 
obtaining quantitative information on the structural, electronic and 
vibrational properties of these molecules. Our results 
suggest that ${\rm C}_{48}{\rm N}_{12}$ has potential applications 
as semiconductor components, nonlinear optical materials, and possible 
building blocks for molecular electronics and photonic devices. 

\end{abstract}

\begin{multicols}{2} 

\section{Introduction}

Graphite is a stable and abundant solid 
form of pure carbon\cite{benedek}. In this form, three valence electrons 
of each carbon atom form three strong $sp^2$ trigonal bonds to three nearest neighbors 
with an equal distance of 0.142 nm, while the fourth valence electron from 
different carbon atoms interacts by weak $\pi$ bonds perpendicular 
to successive sheets with an inter-plane distance of 0.34 nm\cite{aal02}. Diamond 
is another slightly less stable and less abundant crystallographic form of pure carbon 
\cite{benedek}, where each carbon atom is covalently bonded to four neighbors 
via $sp^3$ hybridization at the apexes of a regular tetrahedron. In 1985, a fascinating molecule, 
named ${\rm C}_{60}$ (a truncated icosahedron with 20 hexagonal and 12 
pentagonal faces, and 60 vertices, each of which is at the 
intersection of two hexagonal and one pentagonal faces) was discovered 
by Kroto {\sl et al.}\cite{kroto}, and a new form of pure carbon, 
named fullerenes\cite{curl91}, was born. 

Fullerenes can crystallize in a variety of three-dimensional 
structures\cite{wk90a,drh91}, being made from an even number of 
three-coordinated ${\rm sp}^{2}$ carbon atoms that arrange themselves 
into 12 pentagonal faces and any number ($> 1$) of hexagonal 
faces\cite{curl91}. The macroscopic synthesis of  soot\cite{wk90a}, 
which contains ${\rm C}_{60}$ and other fullerenes in large compounds, 
plus the straightforward purification techniques of the soot which  
make the pure fullerene materials available, have led to extensive studies 
of fullerenes\cite{book1,book3,book4,book4b,book5}.  

Doped fullerenes have also  attracted a great deal of 
interest due to their remarkable structural, electronic, 
optical and magnetic properties\cite{book1,book3,book4,book4b,book5}. 
For example,  the doped fullerenes can 
exhibit large third-order optical nonlinearities\cite{book4,book4b} 
 and be ideal candidates as photonic devices including all-optical switching, data processing, 
 and eye and sensor protection\cite{book4,book4b}. Another example is 
alkali-doped $C_{60}$ crystals, which can become superconducting
\cite{holczer,on99}, for example,  at a critical temperature 
$T_{c}=30$ K \cite{holczer}. In addition to the endohedral doping (inside 
fullerenes) and  exohedral doping (outside fullerenes), there is another type of doping, 
named {\sl substitutional doping}, where  one or more carbon 
atoms of fullerene are substituted by other atoms\cite{book1,book3,book4,book4b,book4c,book5}, 
 due to the unique  structural and electronic properties of fullerenes. 
Because  boron and nitrogen bracket  carbon in the periodic table, much attention has 
been paid to alternate boron- and/or nitrogen-doped compounds\cite{book1,book3,book4,book4b,book4c,book5}.  
Over the past 10 years, boron and nitrogen atoms have been successfully 
used to replace carbon atoms of ${\rm C}_{60}$ and synthesize many 
kinds of  heterofullerenes, ${\rm C}_{\rm 60-m-n}{\rm N}_{\rm m}{\rm B}_{\rm n}$
\cite{book1,book3,book4,book4b,book5,guo91,hummelen95,andreoni96a,hultman01}. In 1995, a very efficient 
method of synthesizing ${\rm C}_{59}{\rm N}$ was reported
\cite{hummelen95}. This method has led to a number of detailed studies of the 
physical and chemical properties of ${\rm C}_{59}{\rm N}$
\cite{book3,book4,book4b,book5,andreoni96a}. Very recently, ${\rm C}_{60}$ with more than one 
nitrogen atom replacing carbon atoms in the cage has been synthesized by  
Hultman {\sl et al.}\cite{hultman01}, and the existence of a novel 
${\rm C}_{48}{\rm N}_{12}$ aza-fullerene\cite{hultman01,stafstrom,rhxie02a} 
was reported. Hence, it would be interesting 
and useful to investigate and predict the structural, electronic, 
vibrational and magnetic properties of this aza-fullerene by performing detailed 
first-principles  calculations. This forms main purpose of the present paper. 

Fullerenes have been challenging molecules for first-principles calculations  because of 
their size\cite{jc95,scuseria96}. Recent advances in {\sl ab initio} 
methods and parallel computing have brought a substantial 
improvement in  capabilities for predicting 
the properties of large molecules. The coupled cluster method 
has been used to predict phenomena in ${\rm C}_{20}$\cite{taylor95}. Other 
first-principles methods, which are less
demanding in terms of computation cost than the coupled cluster 
method, have been used for much larger fullerenes and carbon 
nanotubes. For example, ${\rm C}_{60}$
\cite{disch86,luthi87,scuseria13,haser91,kurita93,hrusak93} has been studied 
 with self-consistent field (SCF)  and Moller-Plesset second-order (MP2) theory, 
${\rm C}_{240}$ \cite{bakowies95} and carbon nanotubes\cite{jc02} with 
density functional theory (DFT) \cite{hohenberg,kohn65}, 
and ${\rm C}_{540}$ \cite{scuseria95} with the Hartree-Fock 
(HF) method. 

 ${\rm C}_{60}$ has the highest symmetry $I_{h}$ 
in the point group,   two types of carbon-carbon 
(CC) bonds (one single bond C-C shared by adjacent five- and six-membered carbon 
rings and one double bond C=C  by adjacent six-membered carbon rings), and 
 two kinds of bond angles (one angle between two adjacent C-C bonds, 
and another one  between a C=C bond and a adjacent C-C  bond)\cite{book1}. In section II, 
we perform full geometry optimizations of ${\rm C}_{48}{\rm N}_{12}$ 
as well as ${\rm C}_{60}$ with both DFT and restricted HF (RHF) 
 methods.  It is found that the ${\rm C}_{48}{\rm N}_{12}$ aza-fullerene has several distinguishing 
features: only one nitrogen atom per pentagon, two nitrogen atoms
preferentially sitting  in one hexagon, ${\rm S}_{6}$ symmetry, six
unique nitrogen-carbon (NC) and nine unique CC bonds. The Mulliken charge analysis 
shows that the doped nitrogen atoms in ${\rm C}_{48}{\rm N}_{12}$  exist 
as electron acceptors and three-fourths of the carbon atoms as electron donors. 
Our best  CC bond lengths and radius 
of ${\rm C}_{60}$ calculated with B3LYP/6-31G* are in excellent agreement 
with  experiment. 

Total energy calculations of the optimized 
${\rm C}_{48}{\rm N}_{12}$ and ${\rm C}_{60}$ are discussed 
 in section III. It is found that the highest occupied molecular 
orbital (HOMO) is a doubly degenerate level with  $a_{g}$ symmetry 
and the lowest unoccupied molecular orbital  (LUMO) is a nondegenerate 
level with  $a_{u}$ symmetry. The calculated HOMO-LUMO energy gap of
${\rm C}_{48}{\rm N}_{12}$ is about 1 eV smaller than that of
${\rm C}_{60}$. For both molecules, the total energies 
calculated with STO-3G, 3-21G and
6-31G basis sets differ from the 6-31G* results by about
1.5\%, 0.6\% and 0.05\%, respectively, and the HOMO-LUMO gaps decrease 
about 5 eV due to electron correlations. For ${\rm C}_{60}$, our calculated 
results are in agreement with other groups' calculations, and our best
 HOMO-LUMO energy gap calculated with B3LYP/6-31G* is in agreement 
with experiment. 

When an external electric field is applied to a molecule, its charges 
are redistributed and  dipoles are induced or reoriented\cite{book9}. The relation between the dipole moment 
${\rm\bf P}$ and the applied field ${\rm\bf G}$ can be written as\cite{book4} 
\begin{eqnarray}
{\rm\bf P} &=&{\rm\bf P}_{0}+\alpha {\rm\bf G}+\frac{\beta}{2}
 {\rm\bf G}^{2}+\frac{\gamma}{6} {\rm\bf G}^{3}+ ...\ , 
\end{eqnarray}
where ${\rm\bf G}$ is the electric field, ${\rm\bf P}_{0}$ is the 
permanent dipole moment, $\alpha$ is the dipole polarizability, 
$\beta$ is the first-order hyperpolarizability, and $\gamma$ is the 
second-order hyperpolarizability. The static dipole 
polarizability (SDP) measures the ability of the valence electrons to find 
a configuration  which screens a static external field
\cite{book4}. It has been shown that molecules with many 
delocalized valence electrons should display 
large SDPs\cite{book4,bianchetti}.  The  
first- and second-order  hyperpolarizabilities play a key 
role in the description of nonlinear optical phenomena since 
a time-varying polarization can act as the source of new components 
of the electromagnetic field\cite{book9}.  In section IV, we calculate 
the SDPs and first- and  second-order hyperpolarizabilities of 
${\rm C}_{48}{\rm N}_{12}$ and ${\rm C}_{60}$.   In comparison to 
isolated carbon or nitrogen atoms, we find  an enhancement in the SDP
due to the delocalized $\pi$ electrons in ${\rm C}_{48}{\rm N}_{12}$ and ${\rm C}_{60}$.  
The calculated SDP for $C_{60}$ is in agreement with experiment. 
The average second-order hyperpolarizability of ${\rm C}_{48}{\rm N}_{12}$ is about
55\% larger than that of ${\rm C}_{60}$.

When a material is doped, its mechanical, electronic, magnetic and optical properties  change 
\cite{book1,book4,book4b,book4c}. The ability to control such induced changes 
is vital to progress in material science. Raman and infrared (IR) 
spectroscopic techniques\cite{book8a,book8b} are useful experimental 
tools to investigate how  doping modifies the structural and dynamical 
properties  of the pristine material and to understand the physical 
origin of such induced changes. Over the past 10 years, both techniques 
have been used widely to study the vibrational properties of ${\rm C}_{60}$ 
\cite{ir1,ir2,ir3,ir4,rm1,rm2}, its derivative compounds
\cite{d1,d2,d3,d4,d5,d6,d7,d8,d9,d10,d11,d12}, and (doped) carbon nanotubes
\cite{rao97a,rao97b,bendiab01,ven02,wzhou02,itkis02,kaw02}. 
It has been shown that ${\rm C}_{60}$ has  in total 
46 vibrational modes including 4 IR-active \cite{ir1,ir2,ir3,ir4} 
and 10 Raman-active vibrational modes\cite{rm1,rm2}. 
These studies have offered a good guide to 
the phonon spectrum in the solid state of these materials. 
In section V, we perform a vibrational analysis and calculate the 
infrared (IR) intensities of ${\rm C}_{48}{\rm N}_{12}$. Fifty eight 
IR-active (i.e., 29 doubly-degenerate and 29 non-degenerate modes) and 58 Raman-active 
(i.e., 29 doubly-degenerate unpolarized and 29 non-degenerate polarized) 
frequencies are determined. The best vibrational frequencies and 
IR results for ${\rm C}_{60}$ calculated with B3LYP are 
in excellent agreement with experiment and demonstrate the desirable 
efficiency and accuracy of this theory for obtaining quantitative 
information on the vibrational properties of these materials. 
Comparison with other groups' calculations of ${\rm C}_{60}$ is 
made and discussed. 

High resolution NMR \cite{nmr1,nmr2,nmr3} gives  spectra which can be analyzed to 
yield  parameters such as  the nuclear magnetic shielding $\sigma$ \cite{ramsey50} 
and  the nuclear spin-spin coupling $J$\cite{ramsey53}, which characterize molecular systems and 
structures.  Both $\sigma$ and $J$ are determined by the 
electronic environments of the nuclei involved.  A satisfactory theoretical 
description of the distribution of electrons in a molecule can lead to 
reliable predictions of $\sigma$ and $J$, which have a number of  
applications, such as the identification of the conformation or 
structure of the species present in a given sample. In section VI  
the GIAO (gauge-including atomic orbital)
\cite{london37} and the CSGT (continuous set of gauge transformations) 
\cite{keith93} methods are utilized for calculating the nuclear magnetic shielding tensor $\sigma$ in 
${\rm C}_{48}{\rm N}_{12}$ and ${\rm C}_{60}$ at both the HF 
and DFT levels of theory. Eight $^{13}{\rm C}$ and two $^{15}{\rm N}$ 
NMR spectral signals  are predicted for ${\rm C}_{48}{\rm N}_{12}$. Our best calculated NMR results for 
${\rm C}_{60}$ are in excellent agreement with experiment. 

Finally, we end  in section VII by giving a summary and outlook on the potential applications 
of  ${\rm C}_{48}{\rm N}_{12}$. 
 
\section{Optimized Geometric Structure}

The geometries of both ${\rm C}_{48}{\rm N}_{12}$ and ${\rm C}_{60}$ 
were fully optimized by using the Gaussian 98 program\cite{gaussian,nist}, 
where we have employed both RHF and DFT methods. 
Also we discuss the effects of basis sets by considering STO-3G, 3-21G, 6-31G 
and 6-31G*\cite{slater30,pople69,pople80a,pople72,pople80b,pople84}. 
For the DFT method, we use  the B3LYP hybrid functional\cite{becke93}.

\

\

\centerline{\epsfxsize=4in \epsfbox{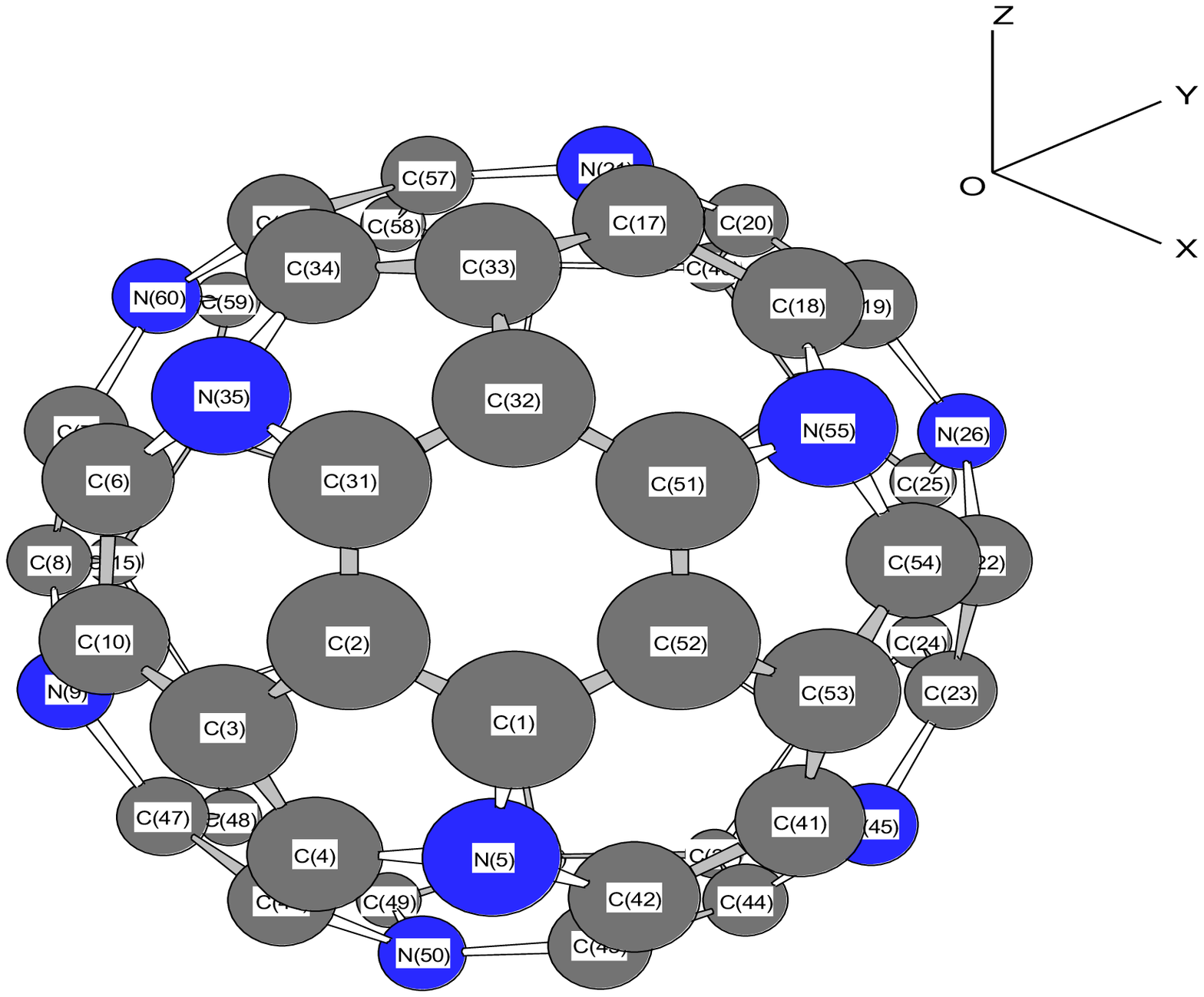}}

\noindent  
{\bf FIG.1}: {\small  Geometric structure of ${\rm C}_{48}{\rm N}_{12}$.  The site numbers \{5, 9, 14, 21, 26, 30, 35, 39, 45, 
50, 55, 60\} are for nitrogen atoms, while the others are for carbon atoms.}
 
\

In Fig.1, we present the geometry of ${\rm C}_{48}{\rm N}_{12}$.  
The {\sl ab initio} calculations show 
that ${\rm C}_{48}{\rm N}_{12}$ has only one nitrogen atom per 
pentagon  and   two nitrogen atoms preferentially sit in one 
hexagon. The symmetry of ${\rm C}_{48}{\rm N}_{12}$ is 
the $S_{6}$ point group\cite{stafstrom,rhxie02a}. The optimized distances 
(or radii) ${\rm R}_{i}$ from the $i$th atom to  the density center of 
the molecule are listed in Table I. We find that there
are 10 unique radii for ${\rm C}_{48}{\rm N}_{12}$, which suggest that
${\rm C}_{48}{\rm N}_{12}$ is an ellipsoidal structure and has 10 unique 
sites (2 for N sites and 8 for C sites), while ${\rm C}_{60}$ has an equal 
radius for each carbon atom.  
As shown in section VI, the 10 unique sites of ${\rm C}_{48}{\rm N}_{12}$ 
can be identified by NMR experiments. Comparing 
the B3LYP results with the RHF results shows that the radii are increased 
by up to 2\% due to the electron correlation. Comparing the 6-31G and 
6-31G*'s results shows that adding polarization functions  decreases 
the  radius of carbon sites but increases the  radius of nitrogen sites. 
In comparison with the results of STO-3G and  
the split valence basis sets, we find that increasing the basis size would 
lead to a decreased radius.  For ${\rm C}_{60}$, the  radius for each carbon site 
calculated by using  B3LYP and the 6-31G* basis set is 3.5502 {\AA}, which is in 
excellent agreement with experiment (R = 3.55 {\AA}) \cite{burgi92}, the LDA 
(local density approximation) calculation (R = 3.537 {\AA}) with a pseudopotential 
approach (PPA)\cite{bohnen95}, and the LDA-based Car-Parrrinello molecular dynamics 
(CPMD) simulation (R = 3.55 {\AA} ) \cite{zhang91}. The success of the B3LYP 
calculation of  the ${\rm C}_{60}$ radius demonstrates the importance of 
electron correlation for an accurate description of a molecular 
geometrical structure. 

The calculated net Mulliken charges ${\rm Q}_{\rm i}$ of
carbon and nitrogen atoms in ${\rm C}_{48}{\rm N}_{12}$
are also listed in Table I. There are two unique types of 
nitrogen atoms in the structure. The net
Mulliken charges ${\rm Q}_{\rm i}/q$ ($q=1.6\times 10^{-19}$)
 on both types of N are negative, for example, $-0.5953$ C and $-0.6002$ C with 
B3LYP/6-31G*, -0.7803 C and -0.7829 C with RHF/6-31G*. The net Mulliken charges of
the carbon atoms in ${\rm C}_{48}{\rm N}_{12}$ separate into two
groups: 1/4 of carbon atoms with  negative
${\rm Q}_{\rm i}$ and the remaining 3/4  with positive ${\rm Q}_{\rm i}$.
Although the Mulliken charge analysis cannot estimate the atomic charges
quantitatively, their signs can be estimated\cite{szabo82}.  From these
results, we find that the doped nitrogen atoms and one-fourth of
the carbon atoms exist as electron acceptors, and  three-fourths
of the carbon atoms as electron donors.
It should be mentioned that we also performed calculations
of net Mulliken charges of carbon and boron atoms in
${\rm C}_{48}{\rm B}_{12}$ \cite{xierh02b}. We found that the doped boron atoms
exist as electron donors and  all carbon atoms as electron acceptors
\cite{xierh02b}. Therefore, ${\rm C}_{48}{\rm N}_{12}$
and ${\rm C}_{48}{\rm B}_{12}$ have opposite
electronic polarizations, while ${\rm C}_{60}$ is isotropic. In the case of
doping into silicon, the V family in the periodic table (for example,
phosphorous) exists as a donor, while the III family (for example, boron)
exists as an electron acceptor. Thus, the B- or N-substituted doping
in ${\rm C}_{60}$ differs greatly from that for silicon. This is
due to the unique structural and  electronic properties  of ${\rm C}_{60}$
\cite{book1}. With respect to the electron correlation and the choice of basis sets, we
find that the absolute value of the net Mulliken charge ${\rm Q}_{\rm i}$ for each atom
in ${\rm C}_{48}{\rm N}_{12}$ increases with an increase of the basis size, but decreases due to the
electron correlation or by adding  polarization functions to a given basis set.

\end{multicols}

\begin{table} 
\noindent
{\bf Table I:}  {\small Net Mulliken charge ${\rm Q}_{\rm i} $ (
$q = 1.6\times 10^{-19}$)  and  radius ${\rm R}_{\rm i}$
( 1 ${\rm\AA}$ = 0.1 nm ) at the site number 
$n_{i}$ in ${\rm C}_{48}{\rm N}_{12}$ 
and ${\rm C}_{60}$ calculated by RHF and B3LYP methods with  
a variety of Pople-style basis sets.}
\begin{center}
\begin{tabular}{cccccccccccc}
 &  &      &         &\multicolumn{2}{c}{STO-3G} 
& \multicolumn{2}{c}{3-21G}  
&\multicolumn{2}{c}{6-31G} & \multicolumn{2}{c}{6-31G*} \\
\cline{5-6}\cline{7-8}\cline{9-10}\cline{11-12}  
Method & Fullerene  & Site Number \{${\rm n}_{\rm i}$\}  
&Atom  & ${\rm R}_{\rm i}$ & ${\rm Q}_{\rm i}/q$ & ${\rm R}_{\rm i}$ & ${\rm Q}_{\rm i}/q$
& ${\rm R}_{\rm i}$ & ${\rm Q}_{\rm i}/q$ & ${\rm R}_{\rm i}$ & ${\rm Q}_{\rm i}/q$  \\
  & &   &  & [${\rm\AA}$] & [C] & [${\rm\AA}$] & [C] & [${\rm\AA}$] & [C] & [${\rm\AA}$] & [C]\\ \hline
RHF & ${\rm C}_{48}{\rm N}_{12}$ & \{ 1, 13, 16, 31, 38, 51\}   & C &3.5452 &0.1073 &3.5059 & 0.3856 & 3.5082 & 0.3546 & 3.4988 & 0.2926  \\     
& &\{ 2, 12, 29, 32, 37, 52\}  & C &3.5591 &-0.0238 &3.5299 & -0.0897 & 3.5271 & -0.0515 & 3.5162 & -0.0621\\
& &\{  3, 11, 28, 33, 36, 53\} & C &3.5602 &-0.0192 &3.5353 & -0.0609 & 3.5275 & -0.0344 & 3.5389 & -0.0480\\
& & \{ 4, 15, 27, 34, 40, 54\}  & C &3.5331 &0.0942 &3.4954 & 0.3905 & 3.4999 & 0.3701 & 3.4716 & 0.3167 \\  
& &\{ 5, 14, 30, 35, 39, 55\}  & N &3.6974 &-0.2406 & 3.5691 &-0.9541 & 3.5685 & -0.9552 & 3.6151 & -0.7803\\ 
& &\{ 6, 18, 24, 42, 48, 58\}  & C &3.4292 &0.0601 &3.4229 & 0.2907 & 3.4299 & 0.2758 & 3.3880 & 0.2209\\   
& &\{ 7, 19, 23, 43, 47, 57\}  & C &3.4023 &0.0893 &3.3874 & 0.3363 & 3.3983 & 0.3058 & 3.3409 & 0.2962\\
& &\{ 8, 20, 22, 44, 46, 56\}  & C &3.4723 &0.0718 & 3.4599 & 0.3191 & 3.4577 & 0.3522 &3.4416 & 0.2433 \\
& &\{ 9, 21, 26, 45, 50, 60\}   & N &3.5870 &-0.2344 & 3.4985 & -0.9878 & 3.4951 & -0.9913& 3.5082 & -0.7829\\ 
& &\{ 10, 17, 25, 41, 49, 59\} & C &3.4851 &0.0952 &3.4586 & 0.3703 & 3.4631 & 0.3737 & 3.4222 & 0.3036\\ 
 &${\rm C}_{60}$ & \{1, 2, 3, 4, 5, 6, ..., 60\} &C&3.5473 & 0 &3.5238 & 0             &3.5300 &0 &3.5226 &0 \\ 
\multicolumn{12}{c}{ } \\
B3LYP&${\rm C}_{48}{\rm N}_{12}$  & \{ 1, 13, 16, 31, 38, 51\}   & C &3.5871 &0.0827 &3.5302 & 0.2770 & 3.5365 & 0.2147 & 3.5171 & 0.1961  \\
& &\{ 2, 12, 29, 32, 37, 52\}  & C &3.5923 &-0.0188 &3.5469 & -0.0422 & 3.5458 & -0.0175 & 3.5275 & -0.0125\\ 
& &\{  3, 11, 28, 33, 36, 53\} & C &3.6019 &-0.0193 &3.5533 & -0.0358 & 3.5492 & -0.0173 & 3.5395 & -0.0298\\ 
& & \{ 4, 15, 27, 34, 40, 54\}  & C &3.5951 &0.0682 &3.5381 & 0.2861 & 3.5408 & 0.2486 & 3.5190 & 0.2266 \\ 
& &\{ 5, 14, 30, 35, 39, 55\}  & N &3.7175 &-0.1883 & 3.5918 &-0.7543 & 3.5945 & -0.6690 & 3.6187 & -0.5953\\ 
& &\{ 6, 18, 24, 42, 48, 58\}  & C &3.5092 &0.0550 &3.4592 & 0.2522 & 3.4706 & 0.1985 & 3.4348 & 0.1919\\ 
& &\{ 7, 19, 23, 43, 47, 57\}  & C &3.4882 &0.0679 &3.4328 & 0.2635 & 3.4465 & 0.2103 & 3.4049 & 0.2124\\ 
& &\{ 8, 20, 22, 44, 46, 56\}  & C &3.5431 &0.0618 &3.4917 & 0.2637 & 3.4946 & 0.2623 & 3.4720 & 0.1998 \\
& &\{ 9, 21, 26, 45, 50, 60\}   & N &3.6302 &-0.1814 & 3.5336 &-0.7787 & 3.5304 & -0.7047 & 3.5335 & -0.6002\\ 
& &\{ 10, 17, 25, 41, 49, 59\} & C &3.5554 &0.0723 &3.5056 & 0.2686 & 3.5087 & 0.2548 & 3.4807 & 0.2110\\
 &${\rm C}_{60}$ & \{1, 2, 3, 4, 5, 6, ..., 60\} &C&3.6034 & 0 &3.5555 & 0             &3.5615 &0 &3.5502 &0 \\ 
\end{tabular}
\end{center}
\end{table}

\begin{table}
\noindent
{\bf Table II:}  {\small Bond lengths (${\rm L}$, 1 ${\rm\AA}$ = 0.1 nm ) 
in ${\rm C}_{48}{\rm N}_{12}$ and ${\rm C}_{60}$ calculated by using B3LYP methods
with a variety of Pople-style basis sets, where $(n_{i},n_{j})$ denotes the site number pair that forms 
a bond.}
\begin{center}
\begin{tabular}{ccccccccccc}
 &      &    & \multicolumn{2}{c}{STO-3G}& \multicolumn{2}{c}{3-21G}
                  &\multicolumn{2}{c}{6-31G} & \multicolumn{2}{c}{6-31G*} \\
\cline{4-5}\cline{6-7}\cline{8-9}\cline{10-11}  
Fullerene&Bond &   $(n_{i},n_{j})$  &   ${\rm L}_{dft}$  &  ${\rm L}_{rhf}$
&  ${\rm L}_{dft}$  & ${\rm L}_{rhf}$ &  ${\rm L}_{dft}$  & ${\rm L}_{rhf}$
& ${\rm L}_{dft}$  & ${\rm L}_{rhf}$ \\
 &     &  &[${\rm\AA}$] &[${\rm\AA}$] &[${\rm\AA}$] &[${\rm\AA}$] &[${\rm\AA}$] &[${\rm\AA}$] &[${\rm\AA}$] &[${\rm\AA}$]\\ \hline
${\rm C}_{48}{\rm N}_{12}$ & CC &(1 ,     2) (12,    13) (16,    29)  &1.4275  &1.3914  &1.4103 &1.3855  & 1.4125 &1.3880  &1.4061   & 1.3836  \\
& & (31,    32) (37,    38) (51,    52)  & &  &  &  &  & & &  \\             
& CC  &(1 ,    52) (2 ,   31) (12,    38) &1.4371   &1.4151 &1.4171 &1.4083  & 1.4216 &1.4134 &1.4155  & 1.4024\\
 & & (13,    29) (16,    37) (32,    51)  &  &  &    &  & & & & \\
& NC & (1 , 5) (13,    14) (16,    30)  &1.4742  &1.4601 &1.432  &1.4206 & 1.4315 &1.4164 &1.4300  &1.4272 \\
& & (31,    35) (38,    39) (51,    55)   &   &  &  &  &  & &  & \\
& CC & (2,     3) (11,    12) (28,    29)  &1.4701 &1.4662  &1.4517 &1.4524 & 1.4488 &1.4478&1.4455 &1.4521 \\
& &(32,    33) (36,    37) (52,    53)  &  &  &  &  &  & & &  \\
& CC & (3,     4) (11,    15) (27,    28) &1.4092  &1.3590  &1.3947 &1.3620  & 1.3971 &1.3652  &1.3901  & 1.3588  \\
 &   & (33,    34) (36,    40) (53,    54) &  &  &  &  &  & & &  \\
& CC  &(3,    10) (11,    59) (17,    33)  &1.4559  &1.4566  &1.4320 &1.4350 & 1.4346  &1.4367 &1.4314   &1.4368  \\
& & (25,    36) (28,    49) (41,    53)   &  &  &  &  & & & & \\
& NC & (4 ,    5) (14,    15) (27,    30)  &1.4707  &1.4553  &1.4307  &1.4128  & 1.4286 &1.4108  &1.4224  &1.4047 \\
& &(39,    40) (34,    35) (54,    55)  &   &  &  &  &  & & &  \\
& CC & (4,    46) ( 8,    15) (20,    40) &1.4567 &1.4589  &1.4326 &1.4367 & 1.4354 & 1.4388 &1.4313 & 1.4372\\
& & (22,    54) (27,    44) (34,    56)   &   &  &  &  & & & & \\
& NC & (5,    42) ( 6,    35) (14,    48) &1.4749  &1.4637  &1.4299  &1.4294 & 1.4317  &1.4278  &1.4287  &1.4310 \\
& &(18,    55) (24,  30) (39,    58)   &   &  &  &  &  & & &  \\
& CC  &(6 ,    7) (18,    19) (23,    24) &1.4399 &1.4352  &1.4211 &1.4182 & 1.4217 &1.4185  &1.4136 &  1.4143\\
 & & (42,    43) (47,    48) (57,    58)  & &  &  &  & & & & \\
&  CC & (6 ,   10) (17,    18) (24,    25)  &1.4119   &1.3641  &1.3963 &1.3664 &1.4004 &1.3712  &1.3941   &1.3634  \\
 &  &(58,    59) (41,    42) (48,    49)  &  &  &  &  & & & & \\
 & CC  &(7 ,    8 ) (19,    20) (22,    23)  &1.4195  &1.3705  &1.4021 & 1.3718 & 1.4068  &1.3761 &1.4021   &1.3703  \\
 &  & (43,    44)  (46,    47) (56,    57)  &  &  &  &  &  & & &  \\
& NC  &(7,    60) (9,    47) (19,    26) &1.4586 &1.4419 &1.4159  &1.4047 & 1.4189 &1.4071 &1.4099 & 1.3958 \\
& & (21,    57) (23,    45) (43,    50)  &   &  &  &  &  & & & \\
& NC & (8 ,    9)  (20,    21) (22,    26) &1.4510 &1.4313  &1.4182 &1.4083  & 1.4149 &1.4012 &1.4084 &1.4056 \\
& &(44,    45) (46,    50) (56,    60)  &  &  &  &  &  & & &  \\
& NC & (9 ,   10) (17,    21) (25,    26) &1.4612  &1.4360  &1.4271 & 1.4043  & 1.4226 &1.3997 &1.4134  &1.3919 \\
& & (41,    45) (49,    50) (59,    60)  &   &  &  &  & & & & \\
${\rm C}_{60}$ & C=C & (1, 52) (2, 31) (3, 10) ... &1.4130  &1.3759  &1.3899 &1.3671 & 1.3981  &1.3750 &1.3949    &1.3732  \\
 & C-C & (1, 2) (1, 5) (2, 3) ... &1.4773  &1.4628 &1.4601 &1.4529 & 1.4592  &1.4524 &1.4539    &1.4487 \\  
\end{tabular}
\end{center}
\end{table}

\begin{multicols}{2}

The optimized CC and NC bond lengths in ${\rm C}_{48}{\rm N}_{12}$ are
listed in Table II. We find that there are 6 unique NC and 9 unique CC bonds.
In comparison with
the calculated CC bond lengths for ${\rm C}_{60}$ shown in Table II,
the CC bond length in ${\rm C}_{48}{\rm N}_{12}$, in general, is less than the single C-C 
bond length of ${\rm C}_{60}$ due to the redistribution of the electron
density. It is also found that the bond length increases due to the
electron correlation, but decreases as we increase  the basis
size or include  the polarization function.
 
In comparison with experimental data available for ${\rm C}_{60}$
listed in Table III,  we find that the two kinds of bond
lengths of ${\rm C}_{60}$  calculated by using the B3LYP with a
large basis set 6-31G* are in good agreement with the results measured
by X-ray powder diffraction (XRPD)\cite{david91},
NMR\cite{yannoni91,johnson90},
gas-phase electron diffraction (GPED)\cite{hedberg91} or X-ray
crystallography technique (XRCT)\cite{burgi92}.

For comparison, Table III also lists the calculated CC bond lengths
for  ${\rm C}_{60}$ with a selection of previous
theoretical calculations. Given the low computational cost of
H\"{u}ckel theory, the bond lengths\cite{haddon86} predicted by
this theory are remarkably satisfactory. The semi-empirical QCFF/PI
(quantum-chemical-force-fields for $\pi$ electrons)\cite{negri88}  does not
predict as good bond lengths as the H\"{u}ckel theory since it has been
parameterized mainly with respect to frequencies of conjugated hydrocarbons.
The CC bond lengths calculated by using the semi-empirical MNDO (modified
neglect of differential overlap)\cite{mndo88}
and the extended Hubbard model (EHM)\cite{kallay98} are a little improved.
These  theoretical approaches empirically include the effect of electron correlation
found in conjugated $\pi$-systems.
The HF\cite{disch86,luthi87,scuseria13,hrusak93} and
SCFMO (self-consistent field with MO) \cite{kurita93}  calculations
are in agreement with our RHF results.  As listed in Table III, the calculated
MP2 bond distances\cite{haser91} usually decrease by
about 0.01 {\AA} when more $d$ functions are added to the basis set,
demonstrating the necessity of including polarization functions
in calculations with correlation.  Based on the differences
between the HF and MP2 data, it is evident that electron correlation effects
should be considered in an accurate description of the equilibrium structure of
a molecule. The CC bond lengths calculated with
the LDA\cite{xqwang93,dad95,hara01},
LDA-PPA\cite{bohnen95}, and LDA-based
CPMD simulation\cite{zhang91,onida94} are
in agreement with our B3LYP's results and demonstrate the importance of electron
correlation effects in giving accurately the equilibrium structure of
a molecule.

\
 
\noindent
{\bf Table III:}  {\small Equilibrium CC bond lengths (1 {\AA} = 0.1 nm)
 of ${\rm C}_{60}$ from previous theoretical predictions and experimental findings. STO-DZP and
DNP  denote double-zeta STO's and double numerical basis with polarization
functions, respectively.}
\begin{center}
\begin{tabular}{cccc}\hline\hline
 Method &  C-C  & C=C & Reference \\
        & [ ${\rm \AA}$ ] &  [ ${\rm \AA}$ ]  & \\  \hline
H\"{u}ckel & 1.436 & 1.418  &\cite{haddon86} \\
QCFF/PI & 1.471 & 1.411   & \cite{negri88} \\
MNDO     & 1.465 & 1.376  &\cite{mndo88}\\
EHM    & 1.446 & 1.402  &\cite{kallay98}\\
SCFMO & 1.49 & 1.43   &\cite{kurita93} \\
HF/STO-3G & 1.463 & 1.376  &\cite{disch86}\\
HF/3-21G &1.453 & 1.367   & \cite{hrusak93}\\
HF/DZ  & 1.451 & 1.368   & \cite{luthi87} \\
HF/STO-3G & 1.463 & 1.376& \cite{scuseria13}\\
HF/DZ  & 1.451 & 1.368  & \cite{scuseria13} \\
HF/DZP  & 1.450 & 1.375  & \cite{scuseria13} \\
HF/TZP  & 1.448 & 1.370  & \cite{scuseria13} \\
MP2/DZ  & 1.470 & 1.407  & \cite{haser91} \\
MP2/DZP  & 1.451 & 1.412 & \cite{haser91} \\
MP2/TZP  & 1.446 & 1.406 & \cite{haser91} \\
LDA/STO-DZP & 1.436 & 1.384 & \cite{hara01}\\
LDA/DZP & 1.445 & 1.395 & \cite{dad95} \\
LDA/DNP   & 1.444 & 1.391 & \cite{xqwang93}\\
LDA-PPA &1.449 & 1.390  &\cite{bohnen95}\\
LDA-CPMD  & 1.45 & 1.40  &\cite{zhang91} \\
LDA-CPMD  & 1.45 & 1.39 &\cite{onida94} \\
Exp./XRCT &1.4459 & 1.3997  & \cite{burgi92}\\
Exp./XRPD & 1.455 & 1.391  & \cite{david91}\\ 
Exp./NMR & 1.45 & 1.40  & \cite{yannoni91}\\
Exp./NMR & 1.46 & 1.40  & \cite{johnson90}\\
Exp./GPED & 1.458 & 1.401  &\cite{hedberg91} \\ \hline
\end{tabular}
\end{center}

As mentioned before,  ${\rm C}_{60}$ has only two kinds of bond angles\cite{book1}, 
$108^{\rm o}$ (the angle between two adjacent single C-C bonds) and $120^{\rm o}$
(the angle between a double C=C bond and an adjacent single C-C bond).  Fig.2(a-d) 
show the distribution of (C-C-C, C-N-C, C-C-N) bond angles in ${\rm C}_{48}{\rm N}_{12}$ 
calculated by using both RHF and B3LYP methods with several different basis sets. 
One-third and two-thirds of the bond angles in ${\rm C}_{48}{\rm N}_{12}$ fluctuate 
around $108^{\rm o}$  and $120^{\rm o}$, respectively. Comparing Fig.2(a) and 2(b) 
shows that increasing the basis size leads to smaller fluctuations in the bond angle 
distribution  (BAD) around either $108^{\rm o}$ or $120^{\rm o}$. In comparison with Fig.2(b), 
Fig.2(c) exhibits enhanced fluctuations in the BAD as polarization functions are added to the 6-31G 
basis set. Comparing Fig.2(c) with Fig.2(d), we find that the BAD in ${\rm C}_{48}{\rm N}_{12}$ are 
decreased due the effect of electron correlation.

\section{Total Electronic Energy}
 
We performed total energy calculations of ${\rm C}_{48}{\rm N}_{12}$ and
${\rm C}_{60}$  by using both RHF and B3LYP with STO-3G, 3-21G, 6-31G
and 6-31G* basis sets. The results are  summarized in
Table IV. The orbital energies are shown in Fig.3, where the orbital
symmetries are also labeled.
 
Table IV demonstrates the convergence of the total energy calculations of
both RHF and B3LYP methods with respect to the basis sets. For both ${\rm C}_{48}{\rm N}_{12}$ and ${\rm C}_{60}$,
the total energies calculated with STO-3G, 3-21G and 6-31G basis
sets differ from the 6-31G* basis set
results by about 1.5\%, 0.6\% and 0.05\%, respectively.
For both molecules, comparing the B3LYP and  RHF results  shows that  the
HOMO-LUMO energy gap $\Delta$ decreases about 5 eV  and the binding energy
$E_{b}$ per atom increases about 2 eV because of the electron  correlations.
The calculated binding energy $E_{b}$ per atom and  HOMO-LUMO energy gap
$\Delta$ of ${\rm C}_{48}{\rm N}_{12}$ are
about 1 eV smaller than those of ${\rm C}_{60}$.
 
Because of the valency of the doped nitrogen atoms, the electronic
properties of ${\rm C}_{48}{\rm N}_{12}$ and ${\rm C}_{60}$ are significantly different. 
As shown in Fig.3(a) and Fig.3(b), the HOMO for ${\rm C}_{60}$  is
fivefold-degenerate with ${\rm h}_{\rm u}$ symmetry,
the LUMO  is threefold-degenerate with
${\rm t}_{\rm 1u}$ symmetry, and the others are threefold-degenerate
with $t_{1g}$ or $t_{2u}$ symmetry, fourfold-degenerate with
$g_{g}$ or $g_{u}$ symmetry, and fivefold-degenerate with $h_{g}$
symmetry.  We notice from Fig.3 that each energy level in
${\rm C}_{48}{\rm N}_{12}$ splits since  the icosahedral symmetry  of
${\rm C}_{60}$ is lost by the substitutional doping.
For ${\rm C}_{48}{\rm N}_{12}$,  the HOMO is  a doubly
degenerate level of ${\rm a}_{\rm g}$ symmetry, the LUMO is a non-degenerate 
level with ${\rm a}_{\rm u}$ symmetry, and the others are specified in Fig.3.
Considering that ${\rm C}_{48}{\rm N}_{12}$ is isoelectronic with ${\rm C}_{60}^{-12}$,
we find that the filling of the energy levels in ${\rm C}_{48}{\rm N}_{12}$ corresponds to a
complete filling of the ${\rm t}_{\rm 1u}$ and ${\rm t}_{\rm 1g}$ levels of
${\rm C}_{60}$.

\end{multicols}

\begin{center}
\epsfig{file=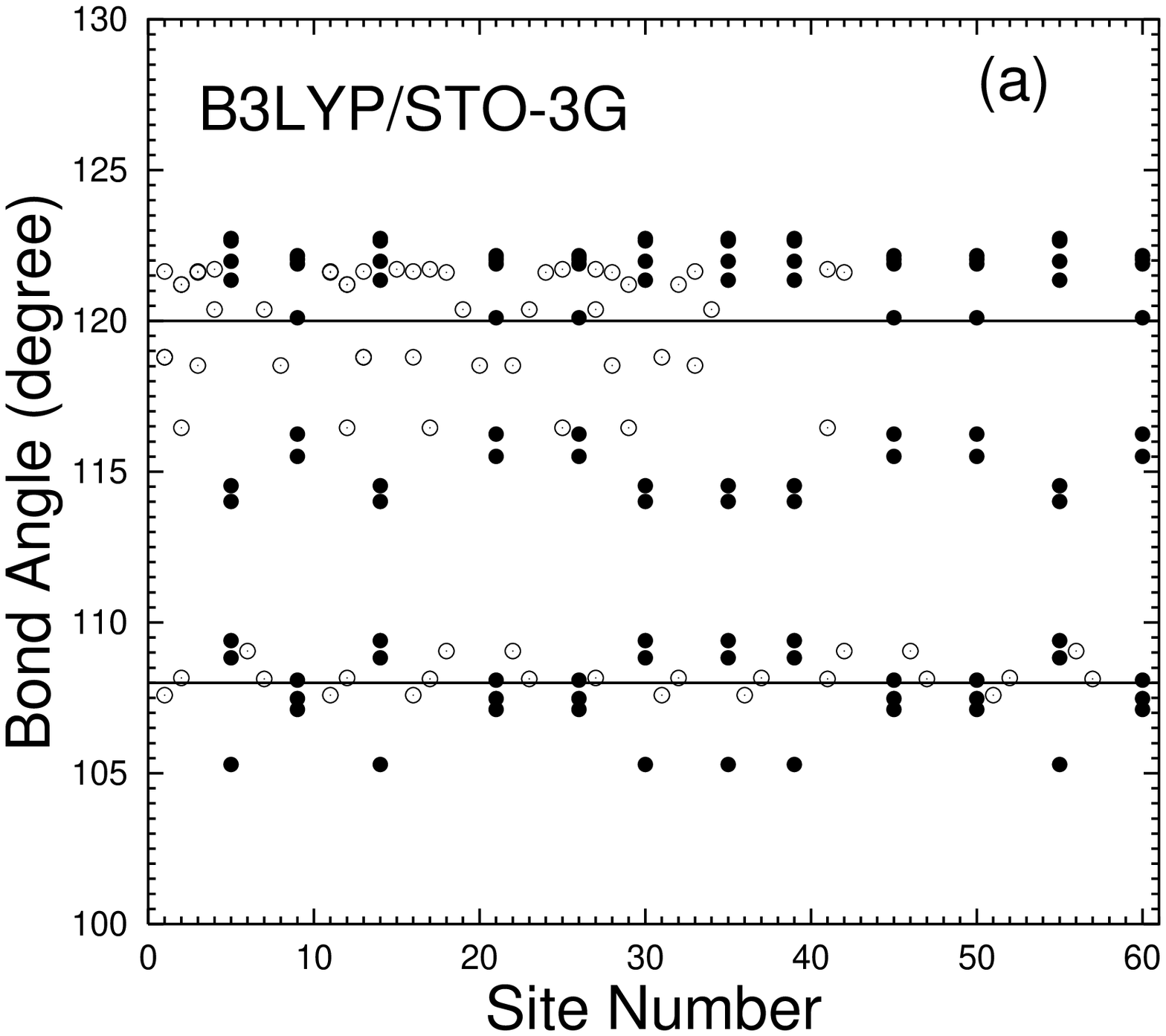,width=5cm,height=5cm}
\hspace{1cm}
\epsfig{file=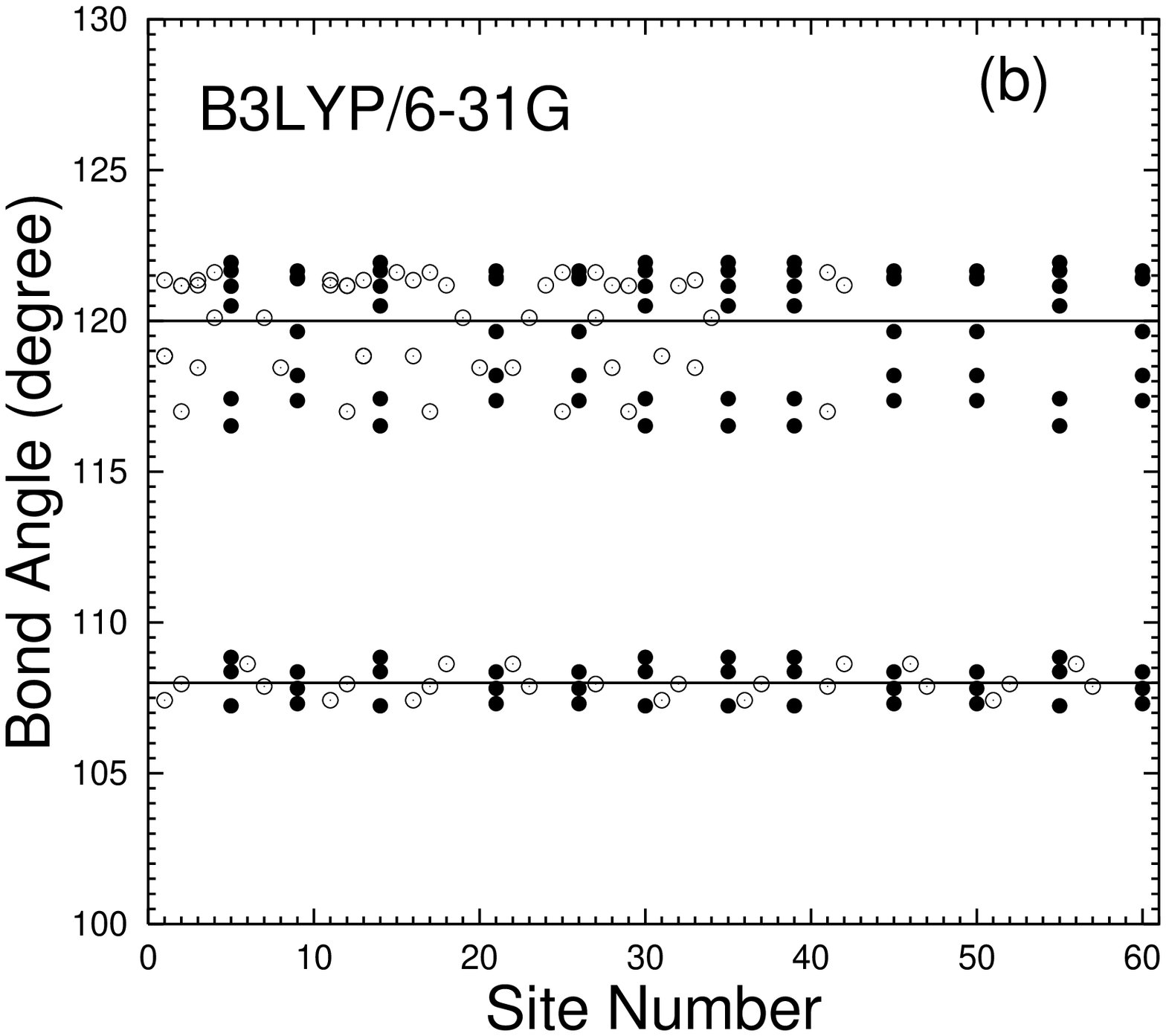,width=5cm,height=5cm}

\epsfig{file=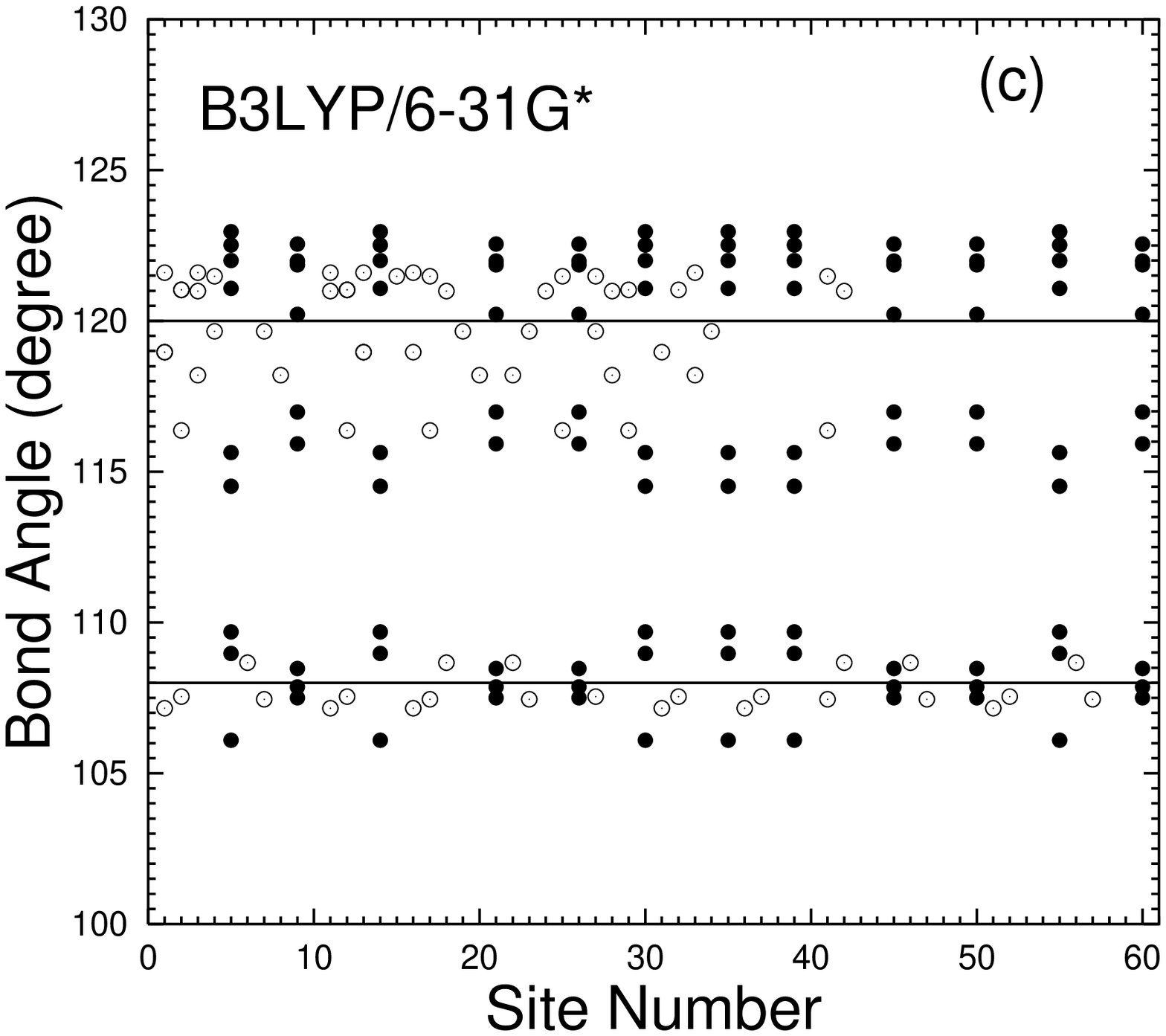,width=5cm,height=5cm}
\hspace{1cm}
\epsfig{file=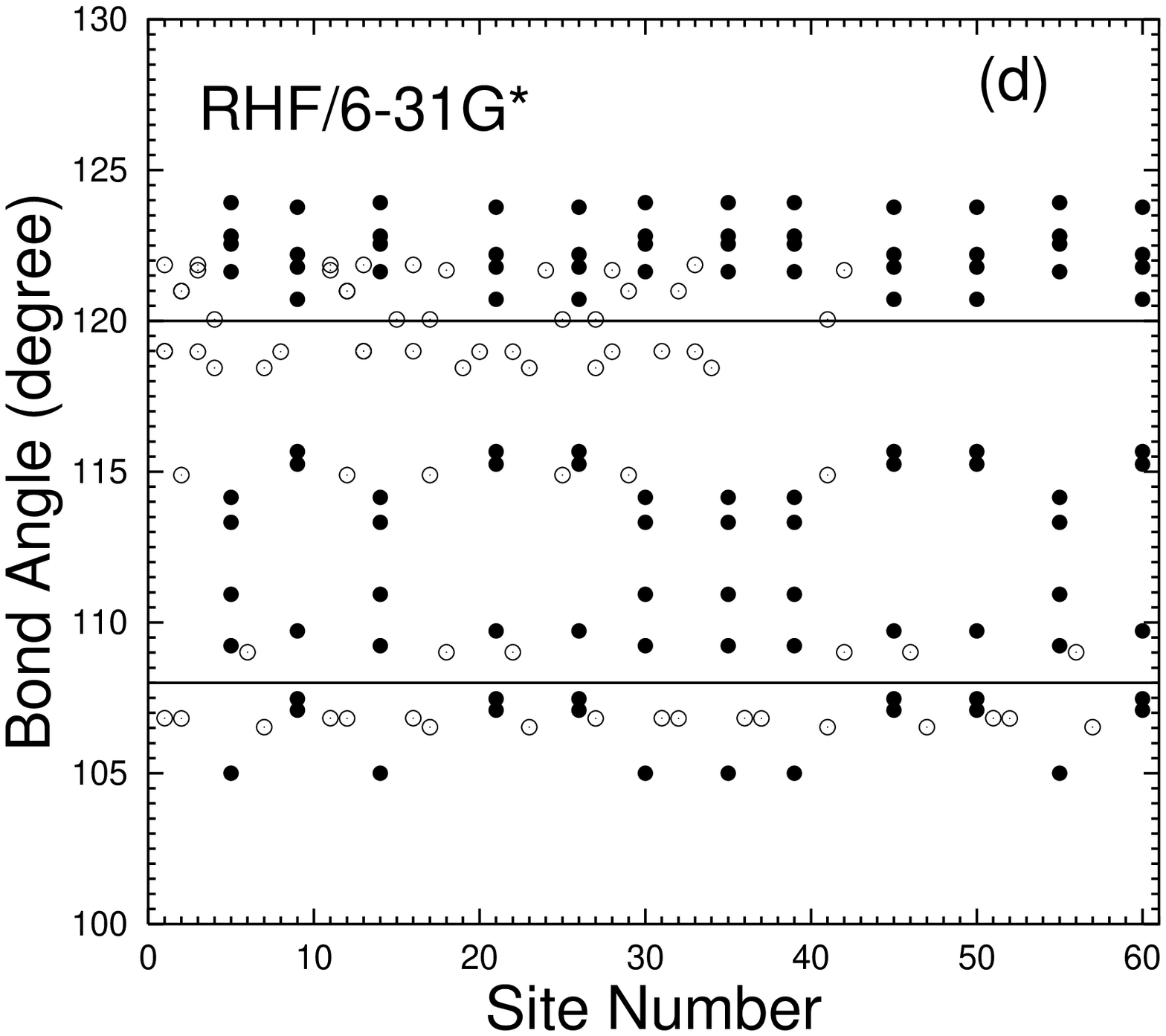,width=5cm,height=5cm}
\end{center}
 
\begin{quote}
{\bf FIG.2}:  {\small {\sl Ab initio} calculations
of C-C-C (open circles) and C-N-C (or C-C-N)  (filled circles)
bond angles in ${\rm C}_{48 }{\rm N}_{12}$: (a) B3LYP/STO-3G; 
 (b) B3LYP/6-31G; (c) B3LYP/6-31G*; (d) RHF/6-31G*. 
The site numbers are labeled as in Fig.1. The solid lines are the 
C-C-C bond angles  (
$108^{o}$ and $120^{o}$) of ${\rm C}_{60}$.}
\end{quote}

\begin{table}
\noindent
{\bf Table IV:} {\small Total electronic energy ($E_{t}$, in eV),
LUMO energy ($E_{lumo}$, in eV), HOMO energy ($E_{homo}$, in eV), 
HOMO-LUMO energy gap ($\Delta$, in eV), binding energy per atom ($E_{b}$, in eV), 
ionization potential ($E_{IP}$, in eV) and 
electron affinity ($E_{EA}$, in eV) of ${\rm C}_{48}{\rm N}_{12}$  and ${\rm C}_{60}$ 
 calculated by using RHF and B3LYP methods with a variety of Pople-style basis sets.}
\begin{center}
\begin{tabular}{llrrrrrrrr}
        &    &\multicolumn{4}{c}{${\rm C}_{60}$} &\multicolumn{4}{c}{${\rm C}_{48}{\rm N}_{12}$} \\
\cline{3-6}\cline{7-10}
Method &Energy & STO-3G  & 3-21G &  6-31G &   6-31G*  &  STO-3G & 3-21G  &  6-31G &   6-31G*     \\  \hline
RHF&$E_{t}$   & -61067.132  &-61471.413 & -61795.062    &-61817.394      & -66397.387& -66846.232  & -67193.956 &-67223.547 \\
& $E_{lumo}$ & 2.688 &-0.632  & -0.709   &-0.117      & 3.797&0.499   & 0.214  &0.643 \\
& $E_{homo}$    & -5.456&-8.323   &-7.919 & -7.644     & -3.287 &-6.506  &-6.203  &-6.189 \\
& $\Delta$       &  8.144      & 7.691 & 7.210 & 7.527  & 7.084  & 7.005  & 6.417 &  6.832 \\
& $E_{b}$       &  5.581      & 4.627 & 4.666 & 4.956  & 4.510   &  3.734    & 3.724  &  4.149  \\  
& $E_{IP}$      &  4.892  & 8.230 & 5.685  & 5.659  & 2.45 & 6.144 & 5.314 & 5.530  \\ 
& $E_{EA}$      &  2.280  & 0.881 & 1.187  & 1.125  & 1.947& 0.146 & 0.209 & 0.267 \\  \\ 
B3LYP & $E_{t}$  &-61446.662   &-61864.864 & -62194.006   & -62209.002   &-66795.835 &-67263.771 &  -67617.031  & -67637.724 \\
 & $E_{lumo}$   &-1.127 &-3.563  &-3.390     &-3.219      & -0.188     &-2.783 &  -2.706   &-2.614   \\
&  $E_{homo}$   &-4.356 &-6.509  &-6.221     &-5.987      & -2.365     &-4.743 & -4.633    & -4.383 \\
& $\Delta$       & 3.229      & 2.946  & 2.831 & 2.768 & 2.177   &  1.96    & 1.927  &  1.774\\
& $E_{b}$  & 7.715 & 6.786 &  6.843 &  6.982   &  6.888  &  6.140 & 6.088  & 6.365 \\  
& $E_{IP}$      &  3.868 &  7.814  & 8.576 & 7.317  &2.488&6.040 & 5.600 &5.663      \\
& $E_{EA}$      &  1.487 &  2.295  & 1.073 & 2.398  &0.804&1.514 & 0.223 &1.493  \\ 
\end{tabular}
\end{center}
\end{table}

\begin{multicols}{2} 
 
In Table IV, we list the calculated ionization potential (IP)  $E_{IP}$
and electron affinity (EA) $E_{EA}$ for both ${\rm C}_{60}$ and
${\rm C}_{48}{\rm N}_{12}$. It shows that ${\rm C}_{48}{\rm N}_{12}$
is a good electron donor, while  ${\rm C}_{60}$  is a good
electron acceptor. The calculated IP for ${\rm C}_{60}$ is in
good agreement with the experiments:
$(7.54\pm 0.01)\ {\rm eV}$ \cite{c60ip1}, $(7.57\pm 0.01)\ {\rm eV}$ \cite{c60ip2},
$(7.58_{-0.02}^{+0.04}\ {\rm eV}$ \cite{c60ip3}, and $(7.59\pm 0.02)\ {\rm eV}$ \cite{c60ip4}.
The calculated EA for ${\rm C}_{60}$ agrees well with the experiments:
$(2.666\pm 0.001)\ {\rm eV}$ \cite{c60ea1},
$(2.689\pm 0.008)\ {\rm eV}$ \cite{c60ea2}.

\begin{center}
\epsfig{file=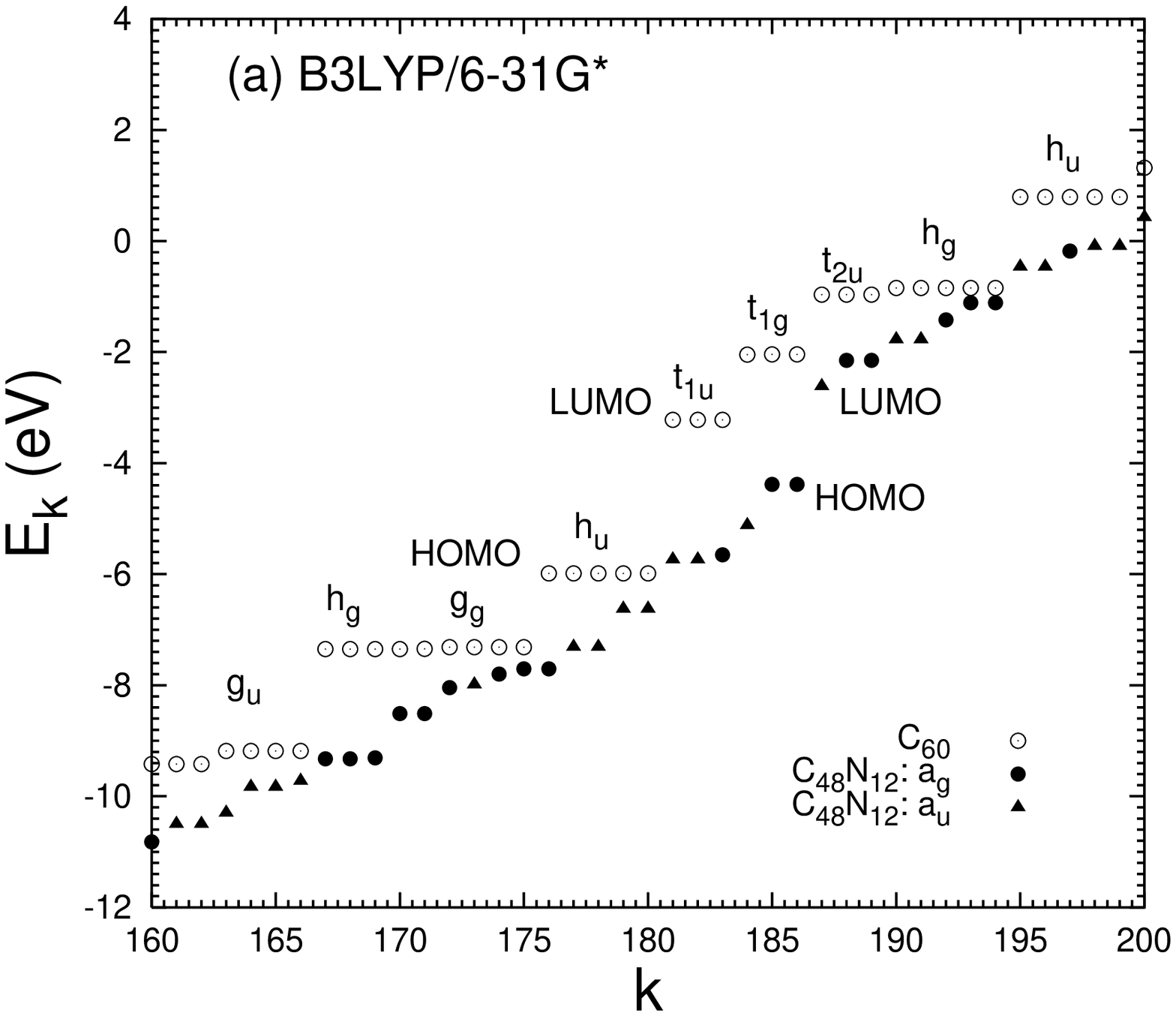,width=5cm,height=5cm}
\hspace{1cm}
\epsfig{file=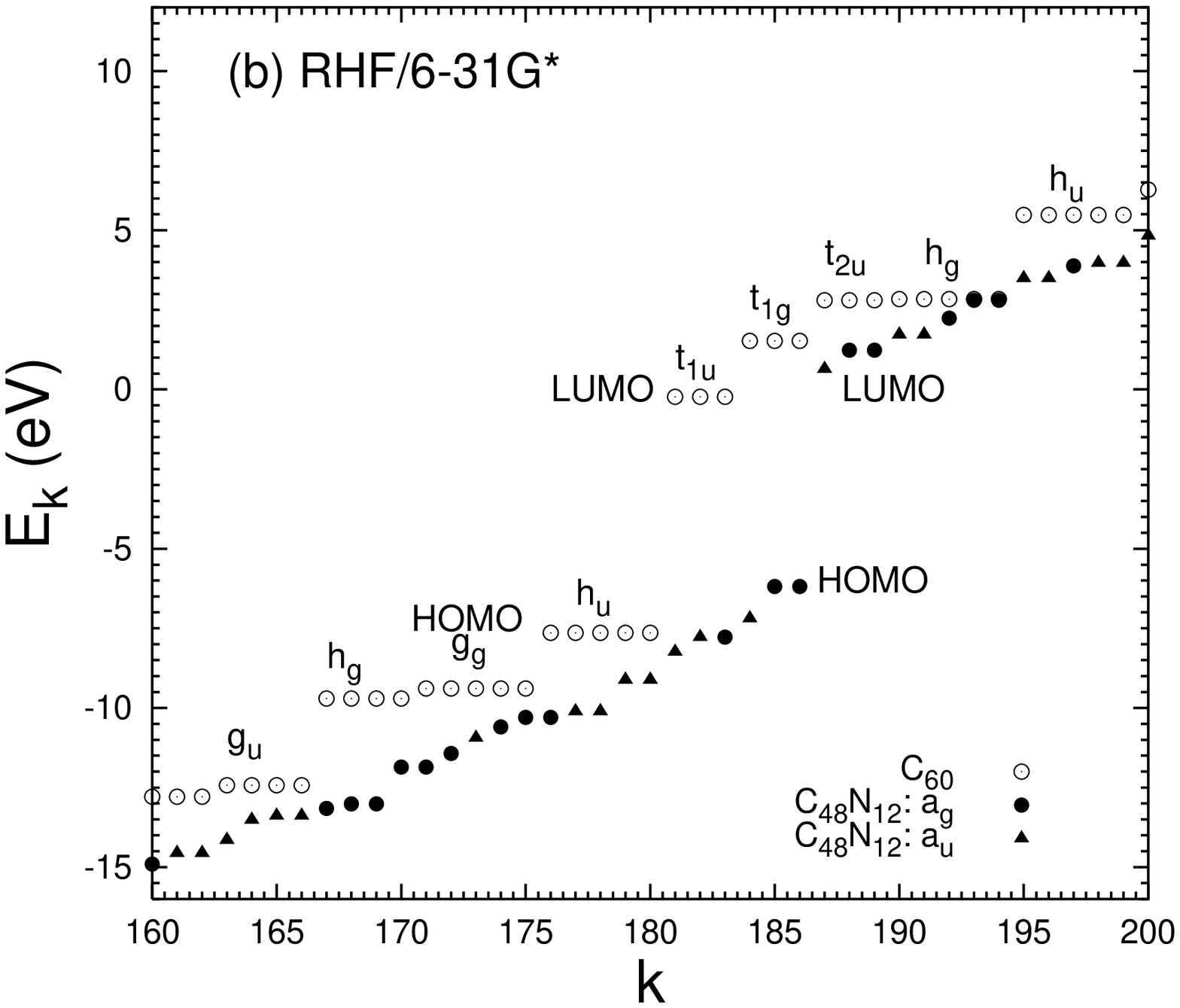,width=5cm,height=5cm}
\end{center}
 
\begin{quote}
{\bf Fig.3}: {\small Orbital energies ${\rm E}_{k}$ of  the
$k$th eigenstate of ${\rm C}_{60}$ and ${\rm C}_{48 }{\rm N}_{12}$ calculated with
(a) B3LYP/6-31G* and (b) RHF/6-31G*. Open circles are for ${\rm C}_{60}$. Filled circles and triangles
are for ${\rm C}_{48 }{\rm N}_{12}$, respectively. The orbital symmetries of energy levels
are shown. }
\end{quote}
  
For comparison,  here we  list the total energies of ${\rm C}_{60}$ for  a 
selection of previous theoretical calculations. Using the HF methods, Scuseria 
\cite{scuseria13} found  that the  total energies of ${\rm C}_{60}$ calculated with 
DZ, DZP and TZP basis sets\cite{disch86,luthi87,scuseria13,hrusak93}
 are -61745.723 eV, -61826.192 eV and -61832.443 eV, respectively. These results  are 
 in agreement with our RHF calculations.  Moreover, using the MP2 method, Scuseria and colleagues 
\cite{haser91} have shown that the total energies of ${\rm C}_{60}$ with 
DZ, DZP and TZP basis sets are -61893.398 eV, -61826.192 eV and -61832.443 eV, 
respectively.  The total energies calculated by using the MP2 method do improve  
the approximation made in HF method. This result demonstrates that electron correlation 
effects cannot be neglected for an accurate prediction of the total energy of a molecule.  

In crystals, the on-site Coulomb interaction between two electrons
 on the same molecule is given by \cite{lof92} 
\begin{eqnarray}
U &=& U_{free} -2{\rm E}_{p}, \\
U_{free} &=& {\rm E}_{IP} - {\rm E}_{EA} - \Delta ,
\end{eqnarray}
where ${\rm E}_{p} = z\alpha e^{2}/(2L^{4})$ ($z$: the number of nearest neighbors, $L$: the distance
between molecules, $\alpha$: the dipole polarizability) is the polarization energy and
$U_{free}$ is the on-site  Coulomb interaction between two electrons 
on the free molecule\cite{book1}. Our plane-wave pseudopotential 
calculations performed by using the CASTEP program \cite{nist,castep} show that
the optimized lattice constants for ${\rm C}_{60}$ and  
${\rm C}_{48}{\rm N}_{12}$-based fcc solids (note: $z = 12$ for both cases) are all around 1.45 nm, while 
the experimental lattice constant = 1.4161 nm for ${\rm C}_{60}$\cite{solid}. Based on 
our first principles results for $E_{IP}$, $E_{EA}$ and $\Delta$ for 
${\rm C}_{60}$ and ${\rm C}_{48}{\rm N}_{12}$ molecules, we arrive at 
the value of the on-site Coulomb interaction,  
${\rm U}  = 1.84, 2.05 \ {\rm eV}$ for ${\rm C}_{60}$ and  ${\rm C}_{48}{\rm N}_{12}$ solids,
respectively. The experimental values for ${\rm C}_{60}$ are 
${\rm U}  =  1.6\pm 0.2$ eV \cite{lof92} and 1.54 eV \cite{schwe98}.   
 Moreover, our DFT/GGA calculations show that
the band gap $E_{gap}$ for the  ${\rm C}_{48}{\rm N}_{12}$ solid 
is about 0.85 eV smaller than that of ${\rm C}_{60}$.
This is consistent with the $\Delta$-U-W relation\cite{book1}  
\begin{equation} 
E_{gap} = \Delta+U-W, 
\end{equation}
where $W$ is the bandwidth for the HOMO- or LUMO-derived energy bands. Assuming the same 
$W$ for both ${\rm C}_{48}{\rm N}_{12}$ and ${\rm C}_{60}$ solids, we 
arrive at 
\begin{equation}
E_{gap}^{C_{48}N_{12}}\approx E_{gap}^{C_{60}}-0.79\ {\rm eV}.
\end{equation} 

The DFT is known to underestimate the band gap of solids. From experiment, 
$E_{gap} = 2.3\pm 0.1$ eV \cite{lof92} and 2.86 eV \cite{schwe98} for the 
${\rm C}_{60}$ solid. Hence, using the approximate relation given by Eq.(5), we 
obtain an estimated band gap for the ${\rm C}_{48}{\rm N}_{12}$ fcc solid of 
$E_{gap}^{C_{48}N_{12}} = 1.7\pm 0.3$ eV. Thus, the ${\rm C}_{48}{\rm N}_{12}$ solid, 
like  ${\rm C}_{60}$, is a  semiconducting material. 

Similarly, we find that the ${\rm C}_{48}{\rm B}_{12}$-based fcc solid (lattice constant $\approx$ 1.45 nm) 
is also a semiconducting material\cite{xierh02b}, 
 having $U = 1.9$ eV, $E_{gap}^{C_{48}B_{12}}-E_{gap}^{C_{60}}=-1.34$  eV, and 
$E_{gap}^{C_{48}B_{12}} = 1.2\pm 0.3$ eV. 

\section{Static (hyper)polarizability}

The static polarizabilities for C$_{48}$N$_{12}$ and C$_{60}$ are presented in 
Table V. The B3LYP and RHF results are obtained  by using the Gaussian 98 
program package \cite{gaussian,nist} and the LDA results by using the ADF (Amsterdam Density 
Functional)  program \cite{nist,ADF:2002,jcc_22_931}. The ADF program uses basis 
sets of Slater functions, where a triple zeta valence basis plus polarization 
is augmented with the field-induced
polarization (FIP) functions of Zeiss et al. \cite{Zeiss:79}. Here this basis set is 
denoted as TZP++ ([6s4p2d1f] for C and N atoms, and [4s2p1d] for H atom) and has recently  been 
used for calculating the second-order hyperpolarizabilities, $\gamma$, for 
C$_{60}$, C$_{58}$N$_{2}$, C$_{58}$B$_{2}$ and C$_{58}$BN \cite{Jensen:02}. 

From the results  listed in Table V, we see that the basis set dependence is identical
for both B3LYP and RHF cases. As expected, improving the basis set increases the
polarizability. Our RHF results for C$_{60}$ are in  good agreement
with a previous study \cite{fowler} which used a similar method and basis sets.
 The B3LYP values are about 10\% larger than the corresponding RHF values.
The LDA results are much larger than both the B3LYP and RHF results. This is
expected since the basis set is larger and expected to predict a more
accurate polarizability \cite{Jensen:02}. Also LDA, in general, predicts
a larger polarizability than does with B3LYP \cite{jcp_109_10180}.
The LDA results for C$_{60}$ are in good agreement with previous  LDA studies
\cite{mrp92,vanGisbergen:97,jcp_115_8773}. We find that the polarizability
of C$_{48}$N$_{12}$ is slightly smaller than the polarizability of C$_{60}$.
For C$_{48}$N$_{12}$
we also find that the $zz$ component is slightly larger than the $xx$ component
except when using RHF and a small basis set. In comparison with the SDP of 
a single carbon or nitrogen atom, we find an enhanced linear polarizability for 
both C$_{60}$ and  C$_{48}$N$_{12}$. This is expected since  both molecules 
have many conjugate $\pi$ electrons delocalized over the entire system.

\end{multicols}
 
\begin{table}
\noindent
{\bf Table V}: {\small Static polarizabilities ($\alpha$, in {\AA}$^3$ = 10$^{-30}$ m$^3$) 
 for C$_{48}$N$_{12}$ and C$_{60}$ with RHF and DFT methods and a variety of basis sets. 
Symmetry relations give for C$_{60}$,  $\alpha_{xx} = \alpha_{yy} = \alpha_{zz}$, and for C$_{48}$N$_{12}$,  $\alpha_{xx} = \alpha_{yy}$.}
\begin{tabular}{lccccccccc}
&\multicolumn{2}{c}{STO-3G}&\multicolumn{2}{c}{3-21G}&\multicolumn{2}{c}{6-31G}
&\multicolumn{2}{c}{6-31G*}&TZP++\\\cline{2-3}\cline{4-5}\cline{6-7}\cline{8-9}\cline{10-10}
&B3LYP&RHF&B3LYP&RHF&B3LYP&RHF&B3LYP&RHF&LDA\\\hline
C$_{60}$&\\
$\alpha_{xx}$&52.5&45.2&64.9&59.5&68.3 &63.2&69.5&64.7&84.7\\
\\
C$_{48}$N$_{12}$\\
$\alpha_{xx}$&51.5&43.5&62.0&55.6&65.6&59.4&66.6&60.0&79.3\\
$\alpha_{zz}$&51.6&41.6&62.6&54.9&66.1&58.5&67.5&60.4&81.5
\end{tabular}
\end{table}

\begin{table}
\noindent
{\bf Table VI}: Static 2nd-order hyperpolarizabilities ($\gamma$, in au. 
 1 au = 6.235378$\times$10$^{-65}$ C$^4$m$^4$J$^{-3}$) 
 for C$_{48}$N$_{12}$ and C$_{60}$  calculated using LDA and TZP++. 
The average 2nd-order hyperpolarizability is given by
$\overline{\gamma} = {1 \over 15}\sum_{i,j}
(\gamma_{iijj}+\gamma_{ijij}+\gamma_{ijji})$.
Symmetry relations give $\gamma_{xxxx}=\gamma_{yyyy}$,
$\gamma_{xxzz} = \gamma_{yyzz}$ and $\gamma_{zzxx}=\gamma_{zzyy}$.
\begin{tabular}{lccccccccc}
&$\gamma_{xxxx}$&$\gamma_{xxyy}$&$\gamma_{zzzz}$&$\gamma_{xxzz}$&$\gamma_{zzxx}$&$\overline{\gamma}$\\\hline
C$_{60}$&137950&45983&137950&45983&45983&137950\\
C$_{48}$N$_{12}$&188780&62880&232970&85120&84790&215222
\end{tabular}
\end{table}

\begin{multicols}{2}

Because ${\rm C}_{60}$ is well separated in a crystal\cite{book1}, one 
can model the fullerene crystal in an electric field as a collection of 
isolated dipoles. For such systems, the Clausius-Mossotti 
relation\cite{cm76} is expected to yield a reasonably accurate relationship 
between the linear polarizability of an isolated molecule and the 
dielectric constant of a fullerene crystal, i.e., 
\begin{equation}
\mid\alpha\mid=\frac{3(\epsilon-1)}{4\pi\rho_{f}(\epsilon+2)},
\end{equation}
where $\rho_{f}$ is the density of fullerene molecules in a 
fcc crystal and $\epsilon$ is the dielectric constant of the fullerene crystal. 
For the purposes of comparison, we note that Hebard {\sl et al.} 
\cite{hebard91} and Ecklund\cite{ecklund92} measured the dielectric
 constants of ${\rm C}_{60}$ films in the  range between 3.9 and 4.0. 
Based on these 
measured dielectric constants,  one can derive an experimental polarizability 
of about $84.9\times 10^{-30}\ {\rm m}^{3}$ for ${\rm C}_{60}$ 
(the density $\rho_{f}=10^{30}/712\ 
m^{-3}$ is given by Quong and Pederson\cite{quong92}).Very recently, using the molecular 
beam deflection technique,  Antoine {\sl et al.}\cite{antoine99} have 
measured the electric polarizability of isolated ${\rm C}_{60}$ molecules 
and obtained a value of $(76.5\pm 8.0)\times 10^{-30}\ {\rm m}^{3}$. Using 
a new optical technique that uses light forces and a time-of-flight 
spectrometer, Ballard {\sl et al.}\cite{ballard00} have made absolute 
measurements of cluster polarizabilities and determined the optical 
polarizability of ${\rm C}_{60}$ at the fundamental wavelength of 
a Nd:YAG laser ( $\lambda = 1064\  {\rm nm}$) to be $(79\pm 4)\times 10^{-30}\ {\rm m}^{3}$. 
The experimental results are in good agreement with our LDA results, 
especially considering that LDA is expected to overestimate the polarizability.

The static first-order hyperpolarizability $\beta$ \cite{book9} of the 
${\rm C}_{48}{\rm N}_{12}$ molecule  is also calculated and found to be 
zero, the same as  that for ${\rm C}_{60}$ molecule. This is 
expected since both ${\rm C}_{48}{\rm N}_{12}$ and ${\rm C}_{60}$ molecule display 
inversion symmetries.  Consequently, this aza-fullerene cannot produce  second-order 
nonlinear optical interactions.

The static 2nd-order hyperpolarizabilities for C$_{48}$N$_{12}$ and C$_{60}$ molecules 
are presented in Table VI. For the calculations of the 2nd-order 
hyperpolarizability, $\gamma$, we use time-dependent (TD) DFT  as described in
Ref.~\cite{Jensen:02,vanGisbergen:97,vanGisbergen:98a,vanGisbergen:98b}. First, 
the 1st-order hyperpolarizability, $\beta$, is calculated analytically in
the presence of a small electric field. Then, the 2nd-order hyperpolarizability
can be obtained by a finite-field differentiation of the
analytically calculated 1st-order hyperpolarizability. For all the TD-DFT calculations we used 
the RESPONSE code \cite{nist,vanGisbergen:98a,vanGisbergen:99} implemented 
in the ADF program \cite{nist,ADF:2002,jcc_22_931}. The small difference between 
$\gamma_{xxzz}$ and $\gamma_{zzxx}$ for C$_{48}$N$_{12}$ is due to 
the numerical method adopted for calculating $\gamma$. 

For the $\gamma$ value of C$_{60}$, we find good agreement with previous first-principles 
results~\cite{vanGisbergen:97,jcp_115_8773,Jonsson:98}. A comparison with 
experiments will not be made for the 2nd-order hyperpolarizability due to 
large differences in the experimental results \cite{book4,book4b,vanGisbergen:97,Jonsson:98}. 
We find that all components of the 2nd-order hyperpolarizability for  C$_{48}$N$_{12}$ 
are larger than for C$_{60}$. This gives an average 2nd-order hyperpolarizability 
of C$_{48}$N$_{12}$ which is about 55 \% larger than the average 2nd-order 
hyperpolarizability of C$_{60}$. The $zzzz$ components of the 2nd-order 
hyperpolarizaility  of C$_{48}$N$_{12}$  is also larger than that found in 
the donor/acceptor substituted C$_{58}$BN molecule~\cite{Jensen:02}. 
 
\section{Vibrational Frequency Analysis}
 
\subsection{Theory}
 
To  lowest order, IR intensities are proportional 
to the derivatives of the dipole moment with respect to the vibrational
normal modes of the material, evaluated at the equilibrium geometry. In detail, 
the IR intensity of the $q$th vibrational mode is given by \cite{ir16} 
\begin{equation}
I_{IR}^{(q)} =\frac{\rho_{p}\pi}{3c}
\left|\frac{d {\bf P}}{d \Xi_{q}}\right|^{2},
\end{equation}
where $\rho_{p}$ is the particle density, $\Xi_{q}$ is the normal
coordinate cooresponding to the $q$th mode and $c$ is the velocity of light.
Since $|d {\bf P}/d \Xi_{q}|$ is the only molecular
property entering the formula, it is often referred to as absolute IR intensity.

To obtain the IR data, one must compute the derivatives
of the dipole moment  with respect to the normal mode
coordinates. These can be viewed as
directional derivatives in the space of 3N nuclear coordinates and expressed
in terms of derivatives with respect to atomic coordinates, $R_{k}$. For the 
$i$th component of the dipole moment ${\bf P}$ (i=x,y,z), we have 
\begin{equation}
\frac{d{\bf P}_{i}}{d\Xi_{q}}=\sum_{k=1}^{3N}\frac{\partial {\bf P}_{i}}{\partial R_{k}}\xi_{kq}, 
\end{equation}
where $\xi_{kq}=\partial R_{k}/\partial\Xi_{q}$ is the $k$th
atomic displacement of the $q$th normal mode.
Then, the necessary derivatives can be expressed in terms of the atomic forces
as follows\cite{ir16,ir17}
\begin{eqnarray}
\frac{\partial {\bf P}_{i}}{\partial R_{k}}&=&
-\frac{\partial^{2}E}{\partial G_{i}\partial R_{k}}=
\frac{\partial F_{k}}{\partial G_{i}},
\end{eqnarray}
where $E$ is the total energy, $G_{i}$ is the $i$th component
of an assumed external electric field ${\bf G}$, and
$F_{k}$ is the calculated force on the $k$th atomic coordinate.
 
\subsection{Normal Vibrations in ${\rm C}_{60}$ and ${\rm C}_{48}{\rm N}_{12}$}
 
Using the Gaussian 98 program\cite{gaussian,nist}, we calculated the harmonic 
vibrational frequencies of both ${\rm C}_{60}$ and ${\rm C}_{48}{\rm N}_{12}$
and considered the effects of the basis sets. It should be mentioned that our 
frequencies have not been scaled. 
 
Table VII and VIII summarize the  vibrational frequencies
for ${\rm C}_{60}$ calculated by using RHF and B3LYP methods,
respectively. As shown by Dresselhaus {\sl et al.}\cite{dress92}, there are 46 
vibrational modes for ${\rm C}_{60}$. These modes are classified in even and odd 
parities and in  ten irreducible representatives
of the $I_{h}$ point group: the \{$a_{g}$, $a_{u}$\}, \{$t_{1g}$, $t_{1u}$,
$t_{2g}$, $t_{2u}$\}, \{$g_{g}$, $g_{u}$\}  and \{$h_{g}$, $h_{u}$\} modes
are non-, threefold-,  fourfold- and fivefold-degenerate, respectively.
Tables VII and VIII demonstrate that increasing the basis size improves 
the accuracy of the predicted vibrational frequencies,
but adding polarization functions to the 6-31G basis set only improves slightly 
the accuracy of the vibrational frequencies.
In choosing a basis set for the first-principles calculation, one must make a
compromise between accuracy and CPU time. Our results shows that the minimum calculation 
can be done in about 4 hours of CPU, while the most expensive calculation 
requires 12 days of CPU. Without significant computational cost, one can do B3LYP/STO-3G
calculation and still obtain results more accurate than any RHF calculations. Going
beyond STO-3G for B3LYP calculations requires a drastic increase in CPU time. Suprisingly,
going just to 3-21G provides the most accurate results, while for the bigger basis
set 6-31G, the results are worse and  adding a polarized function to 6-31G only
slightly improves the results. The 3-21G basis set gives  systematically lower
frequencies  than the 6-31G basis set, while the frequencies obtained from the
6-31G* basis set typically lie  between the results of the other two basis sets.
In contrast, as discussed in section II, 6-31G* does provide the most accurate bond lengths. 
This suggests that the better accuracy of 3-21G is foirtuitous. Increasing the basis set 
to 6-31G stiffens the bonds, while adding the polarization function compensates by softening 
the bonds. In comparison with the B3LYP results,  RHF calculated frequencies are too high 
due to an incorrect description of bond dissociation, while B3LYP with large basis sets
(even the minimum basis set STO-3G)  generally gives results in good agreement with
the experiments of Wang {\sl et al.}\cite{ir3} and  Dong {\sl et al.}\cite{d2}. 
This demonstrates the importance of electron correlation in an accurate 
description of the vibrational frequencies.

For comparison, Table VII lists the  vibrational frequencies
of ${\rm C}_{60}$ calculated by using various theories, for example, the semi-empirical
MNDO\cite{mndo88} and  QCFF/PI\cite{negri88} methods.  Of
these,  the QCFF/PI method, which has been  parameterized mainly 
with respect to vibrational frequencies of conjugated and aromatic
hydrocarbons\cite{warshel72}, results in the best results although 
it gives less satisfactory geometry. Such accurate prediction implies 
that the electronic structures of ${\rm C}_{60}$ is not much different from 
other aromatic hydrocarbons\cite{negri88}. H\"{a}ser {\sl et al.}
\cite{haser91} showed that the approximate harmonic frequencies for 
the two $a_{g}$ vibrational modes of ${\rm C}_{60}$  are 1615 ${\rm cm}^{-1}$ and
487 ${\rm cm}^{-1}$ for HF/DZP,  1614 ${\rm cm}^{-1}$ and
483  ${\rm cm}^{-1}$ for HF/TZP,  1614 ${\rm cm}^{-1}$ and
437  ${\rm cm}^{-1}$ for MP2/DZP, and  1586 ${\rm cm}^{-1}$ and
437  ${\rm cm}^{-1}$ for MP2/TZP. Their HF calculations are in
agreement with our RHF/3-21G results. Their MP2 results are more accurate when
obtained with large basis sets, which also demonstrates the importance
of electron correlation in predicting  the vibrational frequencies.
 
In addition, there have been a number of 2nd nearest-neighbor force-constant models (FCMs)
\cite{wu87,cyvin88,weeks89} which have been  used to calculate the phonon
frequencies of ${\rm C}_{60}$. None of them yield good agreement with
the experimental data.  For example, an empirical force field, which
has been parameterized with respect to polycyclic aromatic hydrocarbons,
is used with  H\"{u}ckel theory and predicts  vibrational frequencies
of the two $a_{g}$ modes of 1409 ${\rm cm}^{-1}$ and  388 
${\rm cm}^{-1}$\cite{cyvin88} that are too low. However, the
modified FCM (MFCM) by Jishi {\sl et al.}\cite{fcm92}  considered interactions
up to the third-nearest neighbors, and the calculated results, as shown
in Table VII, are in excellent agreement with the experiments of  Wang
{\sl et al.}\cite{ir3} and Dong {\sl et al.}\cite{d2}.
 
Table VIII also lists the vibrational frequencies of ${\rm C}_{60}$ calculated  by
other DFT methods, for example, LDA-PPA\cite{bohnen95}, LDA \cite{dad95,hara01} and
DFT-LDA-based CPMD simulations\cite{onida94}. In general, those calculated 
results are in good agreement with experiment. Very recently,  
Choi {\sl et al.}\cite{choi00a} have performed B3LYP vibrational calculations 
of ${\rm C}_{60}$ with a 3-21G basis set but involving scaling of the internal
force constants (SIFC)  $\tilde{K}_{ij}^{int}$ by using Pulay's method\cite{pulay}, i.e.,
\begin{equation}
\tilde{K}_{ij}^{scaled} = (s_{i}s_{j})^{1/2}\tilde{K}_{ij}^{int},
\end{equation}
where $\tilde{K}_{ij}^{int}$ is the force constant in internal coordinates ( the Gaussian 98 program
\cite{gaussian} uses this form),  and $s_{i}$ and $s_{j}$ are scaling factors for the $i$th and $j$th
redundant internal coordinates, respectively. They optimized the scaling factors by minimizing
the root-mean-square deviations between the experimental and calculated scaled frequencies. 
Their results are listed in Table IX. Overall, their scaling procedure improves the accuracy 
for the 46 vibrational frequencies of ${\rm C}_{60}$, especially, for the $a_{g}$, $h_{g}$ and $t_{1u}$ 
vibrational modes. 
 
In Table X and XI, we list the vibrational frequencies for
${\rm C}_{48}{\rm N}_{12}$ calculated with RHF and B3LYP methods and
a variety of Pople-style basis sets.  In contrast with ${\rm C}_{60}$, it is found that
there are in total 116 vibrational modes for  ${\rm C}_{48}{\rm N}_{12}$ because of its
lowered symmetry, $S_{6}$. These vibrational modes are classified into 58
doubly-degenerate and 58 nondegenerate modes. Among those vibrational
modes, there are 58 IR-active (listed in Table X)  and 58 Raman-active modes
(listed in Table XI). Table X and XI show that the electron correlation 
or increasing the basis size results in a redshift of the vibrational 
frequencies. This is similar to that of ${\rm C}_{60}$.

\subsection{IR Intensities in ${\rm C}_{48}{\rm N}_{12}$ and ${\rm C}_{60}$ }
 
We  perform  calculations of IR intensities $I_{IR}$  for both
${\rm C}_{48}{\rm N}_{12}$
and ${\rm C}_{60}$  by using the Gaussian 98 program
\cite{gaussian,nist} with RHF and B3LYP methods.
The calculated IR intensities for ${\rm C}_{60}$ at  the corresponding
frequencies  are listed in Table XII, and those for ${\rm C}_{48}{\rm N}_{12}$
are shown in Fig.4.

For ${\rm C}_{60}$, we note that its IR spectrum is very simple. Namely, it
is composed of 4 IR-active vibrational modes with $t_{1u}$ symmetry.
This is a consequence of the  symmetry of the icosahedral group
\cite{dress92}. Carbon clusters of comparable size, but lower symmetry, have
many more IR-active frequencies. For example, the graphitene
isomer of ${\rm C}_{60}$  has $D_{6h}$ symmetry and 20 IR-active
frequencies\cite{mndo88}. Other examples\cite{mndo88}
included ${\rm C}_{54}$, a planar graphite fragment with $D_{6h}$
symmetry and 22 IR-active frequencies, and ${\rm C}_{50}$, a
spheroidal cluster with $D_{5h}$ symmetry and 22 IR-active
frequencies.  From Table XII, we see that the IR intensity of
a given mode decreases due to the electron correlation and
converges with  increasing basis size.  We find that our intensities 
calculated with B3LYP agree reasonably with experimental spectrum \cite{onoe96} obtained by
{\sl in situ} high-resolution FTIR measurement of a ${\rm C}_{60}$ film. 

The IR intensities $I_{IR}$ at the corresponding
vibrational frequencies for ${\rm C}_{48}{\rm N}_{12}$ are presented
in Fig.4(a)(b)(c) for calculations with B3LYP/STO-3G, B3LYP/3-21G and RHF/3-21G, 
respectively. As discussed above, ${\rm C}_{48}{\rm N}_{12}$ has 29 nondegenerate
and 29 doubly-degenerate IR-active vibrational modes. Similar to the case
of ${\rm C}_{60}$, the IR signals separate into two regions, i.e., a high-frequency
( $> 1000\ {\rm cm}^{-1}$) and a low-frequency
( $\le 1000\ {\rm cm}^{-1}$) region. The IR-active frequencies are redshifted by including the 
electron correlation or  increasing  the basis size. The IR
intensities decrease after including electron correlations and
converge with  increasing  basis size.
The strongest IR spectral lines in both low- and high-frequency regions
are the doubly-degenerate modes located, for example,  at 440 ${\rm cm}^{-1}$ and  1310 ${\rm cm}^{-1}$,
respectively, for the B3LYP/3-21G case. Since experimental IR spectroscopic data do not directly 
indicate the specific type of nuclear motion producing each IR peak, we 
do not give here the normal mode information for each vibrational
frequency and the displacements of each nuclei corresponding to each normal
mode. In Fig.5, taking B3LYP/3-21G calculations as an example, we only show the 
vibrational displacements of sites 1 to 5  (4 C sites and 1 N site) for  
the strongest IR spectral signals in the low- and high-frequency regions.  
It is seen that the pentagon structure  for the low-frequency case expands slightly and

\end{multicols}


 \noindent
{\bf Table VII:} {\small Vibrational frequencies ($\nu$, in ${\rm cm}^{-1}$) 
of ${\rm C}_{60}$ calculated with RHF and a variety of Pople-style 
basis sets. Numbers in the parenthesis are 
the relative errors  to the experimental frequencies, $\nu^{exp}$, 
from Wang {\sl et al.}\cite{ir3} and Dong {\sl et al.}\cite{d2}. The approximated 
CPU times for STO-3G, 3-21G, 6-31G and 6-31G* basis sets are 5 hours, 12 hours, 18 hours and 
39 hours, respectively. Results of other theoretical calculations, for example, QCFF/PI\cite{negri88}, 
MNDO\cite{mndo88} and MFCM\cite{fcm92}, are also listed. }
\begin{center}
\begin{tabular}{ccccccccc}\hline\hline
Mode & STO-3G & 3-21G & 6-31G &6-31G* &QCFF/PI & MFCM  &MNDO   & Exp. \\ \hline 
Even Parity & \multicolumn{8}{c}{ \ } \\  
 $a_{g}$ &1684 (14.5\%) & 1604 (9.1\%)  & 1637 (11.3\%) & 1600 (8.9\%) &1442 (1.9\%) & 1468 (0.1\%) & 1667 (13.4\%) & 1470 \\
      & 553 (11.0\%)  & 518 (4.0\%)  & 526 (5.7\%) & 527 (5.7\%)  &513 (3.0\%)  & 492 (1.2\%) & 610  (22.5\%) & 498  \\
 $g_{g}$& 1802 (18.2\%)  & 1667 (9.3\%)  & 1697 (11.3\%)& 1687 (10.6\%) & 1585 (3.9\%) & 1521 (0.3\%) & 1650  (8.2\%) & 1525   \\
      & 1531 (12.9\%)  & 1426 (5.1\%)  & 1450 (6.9\%) &1441 (6.3\%)    & 1450 (6.9\%) & 1375 (1.4\%) & 1404  (3.5\%) & 1356  \\
      & 1203 (11.8\%)  & 1095 (1.8\%) & 1142 (6.2\%) &1132 (5.2\%)  & 1158 (7.6\%) & 1056 (1.9\%)& 1235  (14.8\%)  & 1076  \\
      & 908 (12.6\%)  & 787 (2.3\%)  & 878 (9.0\%) &836 (3.8\%)   & 770  (4.5\%) & 805 (0.1\%)& 856  (6.2\%)  & 806  \\
      & 653 (5.2\%)   & 646 (4.1\%)  & 643 (3.5\%) &619 (0.3\%)   & 614 (1.1\%) & 626 (0.8\%)&579  (6.9\%)  & 621 \\
      & 574 (18.0\%)  & 529 (9.0\%)  & 549 (12.9\%) &537 (10.5\%)      & 476 (2.1\%) & 498 (2.5\%)& 491  (1.0\%)  & 486 \\
 $h_{g}$& 1912 (21.2\%)  & 1772 (12.3\%)  & 1799 (14.0\%)& 1791 (13.5\%) & 1644 (4.2\%) & 1575 (0.2\%)& 1722  (9.1\%) & 1578 \\
      & 1658 (16.2\%) & 1546 (8.4\%)  & 1585 (11.1\%) & 1562 (9.4\%)     & 1465 (2.7\%) & 1401 (1.8\%)& 1596  (11.8\%) & 1427  \\
      & 1482 (18.4\%)  & 1326 (6.0\%)  & 1377 (10.1\%) &1380 (10.3\%)    & 1265 (1.1\%) & 1217 (2.7\%)&1407  (12.5\%)   & 1251    \\
      & 1290 (17.2\%)   &1184 (7.5\%)   & 1208 (9.8\%) &1208 (9.7\%)   & 1154 (4.8\%) & 1102 (0.1\%)&1261  (14.5\%)  & 1101   \\
      & 886 (14.3\%)    & 828 (6.9\%)   &  843 (8.6\%) & 840 (8.4\%)   & 801 (3.4\%) & 788 (1.7\%) &924  (19.2\%) & 775   \\
      & 836 (17.6\%)   &  761 (7.1\%)    & 821 (15.4\%)& 794 (11.7\%)   & 691 (2.8\%) & 708 (0.4\%)&771  (8.4\%)  &  711     \\
      & 509 (17.5\%)   &  475 (9.7\%)   &  496 (14.5\%)& 482 (11.4\%)   & 440 (1.6\%) & 439 (1.4\%)& 447  (3.2\%) & 433      \\
      & 302 (10.7\%)  & 295 (8.0\%)  &  296 (8.3\%)  &289 (5.9\%)     & 258 (5.5\%) & 269 (1.5\%)& 263  (3.7\%) & 273 \\ 
 $t_{1g}$&1505 (10.9\%)  & 1371 (0.9\%) & 1403 (3.3\%)&1404 (3.4\%) & 1398 (2.9\%) & 1346 (0.9\%)  & 1410  (3.8\%) & 1358  \\
       &966 (1.0\%)  & 940 (3.7\%)  & 939 (3.8\%)&916 (6.1\%)  & 975 (0.1\%) & 981 (0.5\%) & 865  (11.4\%) & 976 \\
       & 687 (36.9\%)  & 618 (23.1\%)  & 670 (33.4\%) & 640 (27.5\%) & 597 (18.9\%) & 501 (0.2\%) & 627  (24.9\%) & 502  \\
 $t_{2g}$&1619 (19.1\%)  & 1453 (6.9\%)  & 1504 (10.6\%) & 1511 (11.1\%) & 1470 (8.1\%) & 1351 (0.7\%) & 1483  (9.0\%) & 1360  \\
       & 912 (0.2\%)  & 907  (0.8\%) & 898 (1.8\%)  &868 (5.0\%)  & 890 (2.6\%) & 931 (1.8\%) & 919  (0.5\%) & 914     \\
       & 903 (4.3\%)  & 701 (19.0\%) & 828 (4.3\%) &734 (15.1\%)  & 834 (3.6\%) & 847 (2.1\%)& 784  (9.4\%)  & 865   \\
       & 647 (14.1\%)  & 637 (12.3\%)  & 634 (11.7\%) & 613 (8.1\%) &637 (12.3\%) & 541 (4.6\%) &591  (4.2\%) & 567    \\ \hline
 Odd Parity & \multicolumn{8}{c}{ \ } \\  
 $a_{u}$ & 1111 (2.8\%)  & 1112 (2.5\%)  & 1097 (4.1\%) & 1061 (7.2\%) & 1206 (5.5\%) & 1142 (0.1\%) & 972 (15.0\%)  & 1143 \\
$g_{u}$ & 1701 (17.7\%) & 1562 (8.0\%)  & 1607 (11.1\%) & 1597 (10.4\%)  &1546 (6.9\%) & 1413 (2.3\%)&1587 (9.8\%) & 1446 \\
 & 1527 (16.6\%) & 1406 (7.3\%)  & 1447 (10.5\%) & 1437 (9.7\%)  &1401 (6.9\%) & 1327 (1.3\%) &1436 (9.6\%)  & 1310 \\
 & 1113 (14.7\%)  & 1031 (6.3\%)  & 1052 (8.4\%)  & 1050  (8.3\%)       &1007 (3.8\%) & 961 (0.9\%) &1110 (4.1\%)  & 970 \\
 & 922 (0.2\%)  & 859 (7.0\%)  & 864 (6.5\%) &826 (10.6\%)         &832 (10.0\%) & 929 (0.5\%)&914 (1.1\%) &  924 \\
 & 863 (13.5\%)  & 738 (2.9\%)  & 846 (11.4\%)& 786 (3.4\%)   &816 (7.4\%) & 789 (3.8\%) &750 (1.3\%) & 760\\
 &414 (3.5\%)  & 390 (2.4\%)  & 400 (0.1\%) & 390 (2.4\%)   &358 (10.5\%) & 385 (3.8\%) &362 (9.5\%)  & 400 \\
$h_{u}$  & 1905 (22.2\%)   &1762 (13.0\%)  & 1791 (14.9\%)&1784 (14.5\%)  &1646 (5.6\%) & 1552 (0.4\%)& 1709 (9.6\%) & 1559  \\
  & 1597 (15.3\%)    &1447 (4.5\%)  & 1487 (7.4\%) & 1491 (7.7\%)          &1469 (6.1\%) &1385 (0.0\%)& 1467 (5.9\%) & 1385   \\
  & 1453 (30.0\%)  & 1320 (18.2\%)  & 1353 (21.2\%) & 1354 (21.2\%)         &1269 (13.6\%) & 1129 (1.1\%)&1344 (20.3\%)  &  1117  \\
  & 886 (10.5\%)   &793 (1.0\%)  & 858 (7.1\%) & 824 (2.8\%)              &812  (1.4\%)  & 801 (0.0\%)&822 (2.6\%)  & 801  \\
  & 777 (11.7\%)   & 753 (8.2\%)   & 761 (9.3\%) & 738 (6.0\%)            & 724 (4.0\%) & 700 (0.6\%)& 706 (1.4\%) & 696      \\
  &640 (13.6\%)   & 587 (4.2\%)    & 613 (8.9\%) & 592 (5.2\%)            & 531 (5.7\%) & 543 (3.6\%)&546 (3.0\%)  & 563      \\
  & 463 (35.0\%)   & 455 (32.7\%)      & 454  (32.4\%) & 442 (28.7\%)      & 403 (17.5\%) & 361 (5.0\%)& 403 (17.5\%) & 343     \\ 
  $t_{1u}$ & 1637 (14.5\%)  & 1553 (8.6\%)  & 1587 (11.0\%) & 1549 (8.4\%)  & 1437 (0.6\%) & 1450 (1.5\%) & 1628 (13.9\%)  & 1429 \\
          & 1396 (18.0\%)  & 1245 (5.2\%)  & 1287 (8.8\%) &1297 (9.6\%)  & 1212 (2.5\%) & 1208 (2.1\%)& 1353 (14.4\%) & 1183 \\
  & 656 (13.9\%)  & 614 (6.5\%)  & 623 (8.1\%) & 625 (8.5\%)   & 637 (10.6\%) & 589 (2.3\%)&719 (24.8\%)   & 576 \\
  & 627 (18.9\%)  & 575 (9.1\%)  & 621 (17.8\%) & 599 (13.6\%)   & 544 (3.2\%) & 505 (4.2\%)& 577 (9.5\%)  & 527 \\
  $t_{2u}$ & 1828 (15.9\%)  & 1709 (8.4\%)  & 1728 (9.5\%) &1713 (8.6\%)  & 1558 (1.2\%) & 1575 (0.1\%)& 1687 (7.0\%)  & 1577 \\
  & 1327 (29.3\%)  & 1214 (1.1\%)  & 1259 (4.5\%) & 1257 (4.7\%)   &1241 (3.3\%)& 1212 (0.9\%)&1314 (9.4\%) & 1201 \\
  & 1074 (4.7\%)  & 995 (3.1\%)   & 1025 (0.1\%)  & 1014 (1.2\%)  &999 (2.6\%)& 1025 (0.1\%) &1134 (10.5\%)  & 1026 \\
& 835 (22.8\%)  & 765 (12.6\%)  & 830 (22.1\%) & 799 (17.5\%)   &690 (1.5\%) & 677 (0.4\%) & 776 (14.1\%) & 680 \\
  & 393 (10.5\%)  & 377 (5.9\%)   & 385 (8.0\%)  & 372 (4.5\%)   &350 (1.7\%) & 367 (3.1\%) & 348 (2.2\%) & 356 \\ \hline
\end{tabular}
\end{center}

\newpage 

\noindent
{\bf Table VIII:} {\small Vibrational frequencies $\nu$ (${\rm cm}^{-1}$)
of ${\rm C}_{60}$ calculated by using the B3LYP method with a variety of Pople-style  basis sets. 
Numbers in the parenthesis  are the relative errors of the calculated frequencies to
 the experimental frequencies listed in Table VII.  
The approximated CPU times for STO-3G, 3-21G, 6-31G and 6-31G* basis sets are 4 hours, 5 days, 8 days and
12 days, respectively.
The other theoretical results are from Bohnen {\sl et al.}\cite{bohnen95}, Dixon {\sl et al.} \cite{dad95}, Hara {\sl et al.}\cite{hara01}, 
 and Onida {\sl et al.}\cite{onida94}. }
\begin{center}
\begin{tabular}{ccccccccc}\hline\hline
{\small Mode} & {\small  STO-3G  } & {\small  3-21G  } & {\small  6-31G  } & {\small  6-31G*  } 
& {\small  Hara  } & {\small  Dixon   } & {\small  Onida  } & {\small  Bohnen  } \\ \hline
Even Parity & \multicolumn{8}{c}{ \ } \\  
{\small  $a_{g}$  } & {\small 1549 (5.4\%)  } & {\small  1501 (2.1\%)   } & {\small  1524 (3.7\%) } 
& {\small  1504 (2.3\%)  } & {\small  1531 (4.1\%)  } & {\small  1525 (3.7\%) } 
& {\small  1447 (1.6\%)  } & {\small  1475 (0.3\%)  }   \\
          & {\small  502 (0.9\%)  } & {\small  491  (1.4\%)  } & {\small  496  (0.4\%)  } 
 & {\small  489 (1.8\%)   } & {\small  502 (0.8\%)  } & {\small  499 (0.2\%)  } 
& {\small  482 (3.2\%)  } & {\small  481  (3.4\%)  }  \\
{\small $g_{g}$  } & {\small  1594 (4.5\%)  } & {\small  1524 (0.1\%)  } & 
{\small  1546 (1.4\%) } & {\small  1538 (0.9\%)   } & {\small  1538 (0.9\%)  } 
& {\small  1548 (1.5\%) } & {\small 1479 (3.0\%) } & {\small 1501  (1.6\%)  }    \\
    & {\small  1380 (1.8\%)  } & {\small  1323 (2.4\%)  } & {\small  1342 (1.0\%) } 
& {\small  1334 (1.6\%)   } & {\small  1337 (1.4\%)  } & {\small  1347 (0.7\%) } & 
{\small  1314 (3.1\%) } & {\small 1287  (5.1\%)   }   \\
  & {\small  1127 (4.7\%)  } & {\small  1062 (1.3\%)  } & {\small  1099 (2.1\%) } 
 & {\small  1093 (1.6\%)   } & {\small  1123 4.4(\%)   } & {\small  1122 (4.3\%) } 
 & {\small  1047 (2.7\%) } & {\small 1037  (3.6\%)  }   \\
 & {\small  788 (2.2\%)  } & {\small  690 (14.4\%)  } & {\small  777 (3.7\%)  } 
 & {\small  754 (6.5\%)    } & {\small  759 (5.8\%)  } & {\small  788 (2.2\%) } 
 & {\small  781 (3.1\%) } & {\small  772  (4.2\%)  }   \\
 & {\small  592 (4.6\%)   } & {\small  598 (3.7\%)  } & {\small  594 (4.3\%)  } 
& {\small  577 (7.1\%)    } & {\small  579 (6.8\%)   } & {\small  573 (7.7\%) } 
& {\small 594 (4.3\%) } & {\small  570  (8.2\%)   }   \\
 & {\small  508 (4.6\%)  } & {\small  484 (0.4\%)  } & {\small  500 (2.9\%)  } 
& {\small 489 (0.7\%)     } & {\small  486 (0.0\%)   } & {\small  484 (0.4\%) } 
& {\small 482 (0.8\%) } & {\small  480  (1.2\%)  }  \\
{\small $h_{g}$  } & {\small  1677 (6.3\%)  } & {\small  1609 (2.0\%)  } & {\small  1627 (3.1\%) } 
& {\small  1618 (2.5\%)   } & {\small  1609 (2.0\%)   } & {\small  1618 (2.5\%) } 
& {\small 1573 (0.3\%) } & {\small 1580  (0.1\%)  } \\
         & {\small  1500 (5.1\%)  } & {\small  1436 (0.6\%)  } & {\small  1466 (2.8\%) } 
 & {\small 1455 (1.9\%)    } & {\small  1475 (3.4\%)  } & {\small  1475 (3.4\%) } 
& {\small  1394 (2.3\%) } & {\small 1422  (0.4\%)   }   \\
  & {\small  1332 (6.5\%)  } & {\small  1231 (1.6\%)  } & {\small  1274 (1.8\%) } 
  & {\small 1275 (1.9\%)     } & {\small  1288 (3.0\%)   } & {\small  1297 (3.7\%) } & {\small 1208 (3.4\%) } 
  & {\small 1198  (4.2\%)  }   \\
        & {\small  1166 (5.9\%)   } & {\small  1112 (1.0\%)  } 
 & {\small  1129 (2.6\%) } & {\small 1125 (2.2\%)   } & {\small  1129 (2.5\%)  } 
 & {\small  1128 (2.5\%) } & {\small  1098 (0.3\%) } & {\small 1079  (2.0\%)   }   \\
        & {\small  802 (3.5\%)   } & {\small  781 (0.8\%)  } & {\small   788 (1.7\%) } 
 & {\small 766 (1.2\%)   } & {\small  794 (2.5\%)  } & {\small  788 (1.7\%) } 
& {\small  775 (0.0\%) } & {\small 763  (1.5\%)   }   \\
         & {\small  734 (3.3\%)   } & {\small   678 (4.7\%)   } 
 & {\small  738 (3.8\%) } & {\small 718 (0.9\%)    } & {\small   711 (0.0\%)  } & {\small  727 (2.3\%) } 
  & {\small 730 (2.7\%) } & {\small 716  (0.7\%)   }   \\
         & {\small  449 (3.7\%)  } & {\small  429 (1.0\%)  } & {\small   448 (3.5\%) } 
& {\small 436 (0.8\%)   } & {\small  430 (0.7\%)  } & {\small  431 (0.5\%) } 
& {\small  435 (0.5\%) } & {\small 422  (2.5\%)  }     \\
        & {\small  271 (0.6\%)   } & {\small  271 (0.8\%)   } & {\small   272 (0.5\%) } 
& {\small 266 (2.7\%)    } & {\small  269 (1.5\%)  } & {\small  261 (4.4\%) } & {\small  261 (4.4\%) } 
& {\small  263  (3.7\%)   }    \\ 
{\small $t_{1g}$ } & {\small 1357 (0.1\%)  } & {\small  1278 (5.9\%)  } 
& {\small  1303 (4.0\%) } & {\small 1301 (4.2\%)   } & {\small  1305 (3.9\%)   } 
& {\small  1318 (2.9\%) } & {\small  1284 (5.4\%) } & {\small  1241  (8.6\%)  } \\
        & {\small 865 (11.3\%)  } & {\small  865 (11.4\%)  } & {\small  858 (12.1\%)  } 
& {\small 840 (13.9\%)    } & {\small  842 (13.7\%)  } & {\small  830 (15.0\%)  } 
& {\small 847 (13.2\%) } & {\small  826  (15.4\%)   } \\
    & {\small 594 (18.3\%)  } & {\small  544 (8.3\%)  } & {\small  593 (18.1\%)  } 
& {\small 576 (14.8\%)   } & {\small  565 (12.5\%)  } & {\small  579 (15.3\%)  } 
& {\small  580 (15.5\%) } & {\small  563  (10.8\%)  }   \\
 {\small $t_{2g}$ } & {\small 1431 (5.2\%)  } & {\small  1330 (2.2\%)  } 
& {\small  1372 (0.8\%)  } & {\small 1370 (0.7\%)  } & {\small  1370 (0.7\%)  } 
& {\small  1393 (2.4\%) } & {\small 1257 (7.6\%) } & {\small  1277  (6.1\%)  }  \\
        & {\small  827 (9.5\%)  } & {\small  839 (8.2\%)  } & {\small  829 (9.3\%)  } 
& {\small 804 (12.0\%)    } & {\small  809 (11.5\%)  } & {\small  839 (8.2\%) } 
& {\small  816 (10.7\%) } & {\small  800  (12.5\%)  }   \\
          & {\small  809 (6.4\%)  } & {\small  650 (24.9\%)  } & {\small  766 (11.4\%) } 
& {\small 743 (14.1\%)    } & {\small  765 (11.6\%)  } & {\small  804 (7.1\%) } 
& {\small  789 (8.8\%) } & {\small  788  (8.9\%)   }   \\
          & {\small  581 (2.5\%)  } & {\small  586 (3.3\%)  } & {\small  582 (2.6\%) } 
& {\small  566 (0.1\%)    } & {\small  566 (0.2\%)  } & {\small  551 (2.8\%) } 
& {\small  559 (1.4\%) } & {\small  543  (4.4\%)  } \\ \hline
Odd Parity & \multicolumn{8}{c}{ \ } \\    
{\small  $a_{u}$  } & {\small  994 (13.0\%)  } & {\small  1013 (11.4\%)   } 
& {\small 991 (13.3\%) } & {\small 982 (14.1\%)   } & {\small  968 (15.3\%)  } 
& {\small  972 (15.0\%) } & {\small 934 (18.3\%)  } & {\small 973  (14.9\%)  } \\ 
{\small $g_{u}$  } & {\small  1513 (4.7\%)  } & {\small  1435 (0.8\%)  } & {\small  1470 (1.7\%) } 
& {\small 1461 (1.0\%)   } & {\small  1474 (1.9\%)  } & {\small  1480 (2.4\%) } 
& {\small  1395 (3.5\%) } & {\small 1420  (1.8\%)  } \\ 
         & {\small  1381 (5.4\%)  } & {\small  1303 (0.5\%)  } & {\small  1338 (2.1\%) } 
& {\small 1333 (1.7\%)    } & {\small  1345 (2.7\%)  } & {\small  1359 (3.7\%) } 
& {\small  1289 (1.6\%) } & {\small 1259  (3.9\%)  }  \\ 
    & {\small  1009 (4.1\%)  } & {\small  972 (0.2\%)  } & {\small  985 (1.5\%) } 
& {\small 977 (0.7\%)  } & {\small   989 (2.0\%)  } & {\small  984 (1.4\%)  } 
& {\small  939 (3.2\%) } & {\small 937  (3.4\%)  }  \\ 
    & {\small 815 (11.8\%)  } & {\small 795 (14.0\%)  } & {\small  788 (14.7\%) } 
& {\small 787 (14.8\%)  } & {\small  780 (15.6\%)  } & {\small  830 (10.2\%) } 
& {\small   796 (13.9\%) } & {\small 790  (14.5\%)  } \\ 
    & {\small  784 (3.1\%)  } & {\small  666 (12.4\%)  } & {\small  775 (2.0\%) } 
& {\small 751 (1.2\%)  } & {\small  762 (0.3\%)  } & {\small  762 (0.3\%) } 
& {\small  763 (0.4\%) } & {\small  756 (0.5\%)   } \\ 
    & {\small 368 (8.1\%)  } & {\small  359 (10.3\%)  } & {\small  366 (8.4\%) } 
& {\small 357 (10.7\%)  } & {\small  355 (11.3\%)  } & {\small  350 (12.5\%) } 
& {\small  352 (12.0\%) } & {\small  348  (13.0\%)  }  \\ 
{\small $h_{u}$   } & {\small  1668 (7.0\%)   } & {\small 1596 (2.4\%)  } 
& {\small  1617 (3.7\%) } & {\small 1608 (3.2\%)  } & {\small  1598 (2.5\%)  } 
& {\small  1611 (3.3\%)  } & {\small  1545 (0.9\%) } & {\small 1566  (0.4\%)  }  \\ 
      & {\small  1427 (3.0\%)   } & {\small 1340 (3.2\%)  } & {\small  1371 (1.0\%) } 
& {\small 1369 (11.8\%)   } & {\small  1371 (1.0\%)  } & {\small  1389 (0.3\%) } 
& {\small  1314 (5.1\%) } & {\small 1291  (6.8\%)   }  \\
      & {\small  1295 (16.0\%)  } & {\small  1219 (9.1\%)  } & {\small  1246 (11.6\%) } 
& {\small 1243 (11.3\%)   } & {\small  1243 (11.3\%)  } & {\small  1248 (11.7\%) } 
& {\small  1198 (7.3\%) } & {\small 1175  (5.2\%) }  \\ 
      & {\small  771 (3.7\%)  } & {\small 719 (10.3\%)  } & {\small  763 (4.7\%) } 
& {\small 736 (8.1\%)  } & {\small  742 (7.4\%)  } & {\small  762 (4.9\%)  } 
& {\small 769 (4.0\%) } & {\small 750  (6.4\%)   }  \\  
      & {\small  696 (0.1\%)    } & {\small 674 (3.2\%)   } 
& {\small  697 (0.1\%) } & {\small  679 (2.5\%)   } & {\small  677 (2.7\%)  } 
& {\small  671 (3.6\%) } & {\small  672 (3.4\%) } & {\small  661  (5.0\%)  }  \\ 
      & {\small 564 (0.1\%)    } & {\small  532  (5.5\%)    } & {\small  555 (1.4\%) } 
& {\small 540 (4.1\%)   } & {\small  537 (4.6\%)  } & {\small  541 (3.9\%)  } 
& {\small  540 (4.1\%) } & {\small 527 (6.4\%)  }  \\   
      & {\small  417 (21.5\%)    } & {\small  419 (22.1\%)   } & {\small  418 (21.8\%) } 
& {\small 408 (19.0\%)   } & {\small  411 (19.8\%)  } & {\small  401 (16.9\%) } 
& {\small  404 (17.8\%) } & {\small  388  (13.1\%)  }  \\ 
     {\small    $t_{1u}$  } & {\small  1505 (5.3\%)  } & {\small  1454 (1.7\%)  } 
& {\small  1479 (3.5\%) } & {\small 1464 (2.4\%)   } & {\small  1489 (4.2\%)  } 
& {\small  1486 (4.0\%) } & {\small  1399 (2.1\%) } & {\small 1457 (2.0\%)   }  \\
            & {\small  1266 (7.0\%)  } & {\small  1175 (0.6\%)  } & {\small  1209 (2.2\%) } 
& {\small 1212 (2.5\%)   } & {\small  1222 (3.3\%)  } & {\small  1224 (3.5\%) } 
& {\small  1158 (2.1\%) } & {\small 1143  (3.4\%)  } \\
            & {\small  596 (3.4\%)  } & {\small  582 (0.9\%)  } 
& {\small  587 (1.8\%) } & {\small 570 (1.0\%)   } & {\small  595 (3.3\%)  } 
& {\small  591 (2.6\%) } & {\small  566 (1.7\%) } & {\small 569  (1.2\%)  }  \\
            & {\small  546 (3.6\%)  } & {\small  508 (3.6\%)  } & {\small  553 (4.9\%) } 
& {\small 537 (1.8\%)   } & {\small  528 (0.2\%)   } & {\small  535 (1.5\%) } 
& {\small  541 (2.7\%) } & {\small 514  (2.5\%)   }  \\
{\small  $t_{2u}$  } & {\small  1622 (2.8\%)  } & {\small  1568 (0.6\%)  } 
& {\small  1579 (0.2\%) } & {\small 1568 (0.6\%)   } & {\small  1568 (0.6\%)  } 
& {\small  1571 (0.4\%) } & {\small  1537 (2.5\%) } & {\small  1546  (2.0\%)  } \\
         & {\small  1235 (2.8\%)  } & {\small  1163 (3.2\%)  } & {\small  1201 (0.0\%) } 
& {\small 1199 (0.2\%)   } & {\small  1231 (2.5\%)  } & {\small  1234 (2.7\%) } 
& {\small  1108 (7.7\%) } & {\small 1131  (5.8\%)   }  \\
         & {\small  1002 (2.4\%)  } & {\small  964 (6.0\%)   } & {\small  984 (4.1\%) } 
& {\small 966 (5.9\%)    } & {\small  997 (2.8\%)  } & {\small  996 (2.9\%)  } 
& {\small  936 (8.8\%) } & {\small 945  (7.9\%)  }  \\
         & {\small  734 (7.9\%)  } & {\small  678 (0.3\%)  } & {\small  742 (9.2\%) } 
& {\small 722 (6.1\%)   } & {\small  715 (5.1\%)  } & {\small  726 (6.8\%)  } 
& {\small  774 (13.8\%) } & {\small 725   (6.6\%)  }  \\
         & {\small  354 (0.5\%)  } & {\small  345 (3.1\%)   } & 
{\small  352 (4.3\%) } & {\small 342 (4.0\%)  } & {\small  343 (3.7\%)  } 
& {\small  342 (3.9\%)  } & {\small  340 (4.5\%) } & {\small 343  (3.8\%)   } \\ \hline 
\end{tabular}
\end{center}

\newpage 

\noindent
{\bf Table IX:} {\small  Vibrational frequencies $\nu$ (${\rm cm}^{-1}$)
of ${\rm C}_{60}$ obtained by Choi {\sl et al.}\cite{choi00a} using B3LYP/3-21G but 
involving scaling of the internal force constant by using Pulay's method. 
Numbers in the parenthesis  are the relative errors of the calculated frequencies to
 the experimental frequencies listed in Table VII.} 
\begin{center}
\begin{tabular}{ccccccccccccc}\hline\hline
\multicolumn{6}{c}{Even Parity}& \ \ \  &\multicolumn{6}{c}{Odd Parity}\\ 
\cline{1-6}\cline{8-13}
Mode & $\nu$ & Mode & $\nu$ & Mode & $\nu$ & & Mode & $\nu$ & Mode & $\nu$ & Mode & $\nu$ \\ \hline
$a_{g}$ & 1470 (0.0\%) &         &             &$t_{1g}$& 1290 (5.0\%) &  &$a_{u}$  & 1078 (5.7\%) &          &              &$t_{1u}$ &1433 (0.3\%) \\ 
        & 495 ( 0.6\%) & $h_{g}$ & 1576 (0.1\%)&        & 904 (7.4\%)  &  &         &              &          &              &         &1180 (0.3\%) \\
        &              &         &1427 (0.0\%) &        &  565 (12.5\%)&  &         &              &$h_{u}$   & 1567 (0.5\%) &         & 577  (0.2\%) \\
$g_{g}$ & 1497 (1.8\%) &         &1251 (0.0\%) &        &              &  &  $g_{u}$&1429 (1.2\%)  &          &  1343 (3.0\%)&         & 526  (0.2\%) \\
        & 1348 (0.6\%) &         & 1101 (0.0\%)&        &              &  &         &1315 (0.4\%)  &          & 1214 (8.7\%) & $t_{2u}$& 1524 (3.4\%) \\
        & 1040 (3.3\%) &         &775  (0.0\%) &$t_{2g}$& 1340 (1.5\%) &  &         & 970  (0.0\%) &          & 737 (8.0\%)  &         & 1142 (5.0\%) \\
        &758  (6.0\%)  &         &711  (0.0\%) &        &  831 (9.1\%) &  &         &  797 (13.7\%)&          & 694 (0.3\%)  &         & 955 (6.9\%) \\
        & 592 ( 4.7\%) &         & 431  (0.5\%)&        & 668 (22.8\%) &  &         & 707 (7.0\%)  &          & 535 (5.0\%)  &         & 716 (5.3\%) \\
        & 485 ( 0.2\%) &         & 267  (2.2\%)&        & 614 (8.3\%)  &  &         & 354 (11.5\%) &          & 403 (17.5\%) &         & 340 (4.5\%) \\ \hline  
\end{tabular}
\end{center}

\begin{table} 
\noindent
{\bf Table X:} {\small Fifty eight IR-active frequencies 
($\nu$, in ${\rm cm}^{-1}$) of ${\rm C}_{48}{\rm N}_{12}$ calculated 
by using RHF and B3LYP methods with a variety of Pople-style 
basis sets. } 
\begin{center}
\begin{tabular}{cccccccc|cccccccc}
\multicolumn{8}{c|}{\small Doubly-degenerate Modes} 
&\multicolumn{8}{c}{\small Non-degenerate Modes} \\ 
\cline{2-7}\cline{10-15}
\multicolumn{4}{c}{\small B3LYP} & \multicolumn{4}{c|}{\small RHF} & 
\multicolumn{4}{c}{\small B3LYP} & \multicolumn{4}{c}{\small RHF}   \\
\cline{1-4}\cline{6-8}\cline{9-12}\cline{14-16}
 {\small STO-3G} & {\small 3-21G} & {\small 6-31G} &{\small 6-31G*} & {\small STO-3G} & {\small 3-21G}
 & {\small 6-31G} & {\small 6-31G*}  & {\small STO-3G} & {\small 3-21G} & {\small 6-31G} & {\small 6-31G*}
 & {\small STO-3G} & {\small 3-21G} & {\small 6-31G} & {\small 6-31G*}\\ \hline
{\small  314}   & {\small  308}   & {\small 318} &{\small 309}   &  {\small 358}   &  {\small 338}  &  {\small  350} & {\small344} 
 & {\small 286}   & {\small  287}   &   {\small 297}&{\small 287} & {\small  305}  & {\small  303}   & {\small 317} & {\small 305}\\ 
{\small 363}   &   {\small 359}  & {\small 369 }&{\small 357}  &   {\small  400}   &  {\small 386}  &  {\small    398 }& {\small 381}
 & {\small 328}   &   {\small 318}  & {\small  329}&{\small 325}  & {\small 370}   &  {\small 350}     & {\small 362} & {\small 361} \\ 
{\small 394}   & {\small   404}  & {\small 406}&{\small 395}   &   {\small  430}   & {\small  431}  &  {\small 436} & {\small 419}
&{\small  362}   &   {\small 359}  &  {\small 371} &{\small 360} &  {\small 404}  &  {\small 394}    &{\small 406} & {\small 392} \\
{\small 406}   &  {\small  412}  & {\small 418} &{\small 406}  &    {\small 452}   &  {\small 452}  &  {\small  454}  & {\small 443}
 & {\small 391}   &  {\small  402}  & {\small  406}& {\small 395}  & {\small   428} &  {\small 434}    & {\small 437}  & {\small 423} \\
 {\small 437}   & {\small   422}  & {\small  441} & {\small 425}  & {\small    513}   & {\small  472}  & {\small  495}  & {\small 479}
 & {\small  450}   & {\small   413}  & {\small  447} & {\small 431} & {\small  527}  & {\small  466}    & {\small 498}& {\small 482} \\
{\small 465}   &   {\small 440}  & {\small 464} &{\small 446}  &   {\small   551}  & {\small  497}  &  {\small   523} & {\small 506}
& {\small  470}   & {\small   454}  &  {\small 467}&{\small 460}  &  {\small 556}  &  {\small 505}    & {\small 524}&{\small  522} \\ 
 {\small 496}   & {\small   469}  & {\small 491}&{\small 481}   &  {\small   578}   & {\small  532}  &  {\small  554} & {\small 549}
&{\small  581}   & {\small   567}  & {\small  580}&{\small 579}  & {\small  641}  & {\small  603 }   &{\small 617} &{\small 614} \\
{\small 566}   & {\small   558}  &{\small  569}&{\small 570}   & {\small   619 }   &{\small   591}  &  {\small   606} & {\small 604}
& {\small  592 }  & {\small   600}  &{\small   639}&{\small 611}  &{\small   695}  &{\small   662}     &{\small 720}&{\small 662} \\ 
{\small 623 }  & {\small   585}  &{\small  633} &{\small 606}  & {\small   723 }   & {\small  662}  & {\small     707} &{\small 671}
 &{\small  645}   & {\small   609}  & {\small  653}&{\small 622}  &{\small   735}  &{\small   709}     &{\small 739}&{\small 705} \\
{\small 632}   & {\small   635}  &{\small  662} &{\small 639}  & {\small  738 }   &{\small   730}  & {\small    741} & {\small 719}
 &{\small  656}   & {\small   654}  & {\small  674}&{\small 645}  &{\small   767}  & {\small  734}     &{\small 747}&{\small 722} \\  
{\small 656}   & {\small   665}  &{\small  679} &{\small 658}  &{\small  763 }    &{\small   747}  & {\small     755} &{\small 735}
&{\small  692}   & {\small   698}  & {\small  707}&{\small 686}  &{\small  802}   &{\small   762 }    &{\small 779} &{\small 755} \\ 
{\small 675}   & {\small   686}  &{\small  697}&{\small 675}   &{\small   771 }    &{\small   767}  &  {\small   786} &{\small 757}
 &{\small  735}   &{\small    709}  & {\small  722}&{\small 700}  &{\small  839}   & {\small  800}     &{\small 819} &{\small 770} \\  
{\small 695}   & {\small   701}  &{\small 709} &{\small 695}   & {\small 801  }    &{\small   783}  & {\small    801} &{\small 779}
&{\small  784}   &{\small   798}   & {\small  798}&{\small 778}  &{\small  871}   &{\small   859 }    &{\small 867}&{\small 836} \\  
{\small 722}   & {\small   721}  &{\small  729}&{\small 712}   &{\small   843 }    &{\small   810}  &{\small      829} &{\small 807}
&{\small  798}   &{\small   809}   & {\small  803} &{\small 786} &{\small  899}   &{\small   878 }    &{\small 879}&{\small 853} \\   
{\small 788}   &{\small   804}   &{\small  800} &{\small 782}   & {\small  879 }    &{\small   871}  & {\small    874} &{\small 847}
&{\small  958 }  & {\small  963}   & {\small  968}&{\small 961}  &{\small  1058}  &{\small   1035}    &{\small 1044}&{\small 1030} \\ 
{\small 973}   &{\small   958}   & {\small 965}&{\small 954}   &{\small   1072}    &{\small  1028}  &{\small  1040} & {\small 1020}
&{\small  972}   & {\small  977}   & {\small  982}&{\small 967}  &{\small  1076}  &{\small   1055}    &{\small 1062} &{\small 1045} \\
{\small 1055}  & {\small  1041}  &{\small  1050}&{\small 1040}  &{\small  1162}     &{\small  1118}  & {\small   1135} &{\small 1116}
&{\small  990}   &{\small  1010}   & {\small   993}&{\small 976} &{\small  1095}  &{\small   1102}    &{\small 1092}&{\small 1060} \\ 
{\small 1203}  & {\small  1136}  &{\small  1163}&{\small 1159}  &{\small 1336 }     &{\small  1217}  &{\small   1259} &{\small 1244}
&{\small  1057}  &{\small   1042}  & {\small  1056}&{\small 1047} & {\small  1165} &{\small   1125}    &{\small 1145}&{\small 1124} \\ 
{\small 1231}  & {\small   1172} &  {\small 1199}&{\small 1197} &{\small 1372}    &{\small    1265}  &{\small     1301} &{\small 1292}
 &{\small  1220}  &{\small  1142}  &{\small   1171}&{\small 1177}   &{\small  1351}  &{\small  1232}    &{\small 1272}&{\small 1272}  \\
{\small 1275}  &{\small   1210}  & {\small 1237}&{\small 1237}  &{\small  1415}   &{\small    1304}  & {\small    1341} &{\small 1334}
&{\small  1276}  &{\small   1215}  &{\small   1238}&{\small 1229} & {\small 1393}  &{\small   1311}    &{\small 1345} &{\small 1330} \\
{\small 1348}  &{\small   1280}  & {\small 1306} &{\small 1302} & {\small 1501}   &{\small    1369}  & {\small    1409} &{\small 1397}
 &{\small  1331}  &{\small   1252}  & {\small  1277}&{\small 1270} &{\small  1446}  &{\small   1326}    &{\small 1365}&{\small 1351} \\
{\small 1391}  & {\small  1311}  &{\small  1339} &{\small 1340} &{\small  1552}   &{\small    1413}  & {\small   1454} &{\small 1450}
 & {\small 1362}  &{\small   1293}  &{\small   1318}&{\small 1317} & {\small 1464}  &{\small   1381}    &{\small 1421}&{\small 1377}\\ 
{\small 1440}  & {\small  1354}  &{\small  1387} &{\small 1389} &{\small  1618}   &{\small    1467}  & {\small  1512}  &{\small 1511}
& {\small  1421}  &{\small  1345}   & {\small  1380}&{\small 1384} & {\small 1521}  &{\small   1405}    &{\small 1463}&{\small 1412}\\
{\small 1478}  & {\small  1410}  & {\small 1448}&{\small 1437}  &{\small  1647}   & {\small   1511 } & {\small     1562} &{\small 1540}
 &{\small  1437}  & {\small  1374}  & {\small  1413}&{\small 1394} &{\small  1609}  &{\small   1451}    &{\small 1502}&{\small 1494} \\ 
{\small 1525}  &{\small   1443}  & {\small 1476} &{\small 1473} &{\small  1731}   &{\small  1566 }   & {\small     1613} & {\small 1605} 
 &{\small  1466}  &{\small   1381}  &  {\small  1422}&{\small 1418} &{\small 1621}  & {\small  1490}    &{\small 1532}&{\small 1533} \\ 
{\small 1555}  &{\small    1475} & {\small  1502}&{\small 1499} & {\small 1769}   &{\small  1612}    & {\small      1652}&{\small 1642} 
 &{\small  1506}  &{\small   1418}  & {\small   1458}&{\small 1461} &{\small 1700}  &{\small   1523}    &{\small 1583}&{\small 1574} \\
{\small 1602}  &{\small   1521}  &{\small  1553} &{\small 1553} &{\small 1839}   & {\small  1673}    & {\small   1713} &{\small 1713} 
 &{\small  1560}  &{\small   1497}  & {\small   1529}&{\small 1518} &{\small 1773}  &{\small   1633}    &{\small 1674}&{\small 1662} \\
{\small 1628}  &{\small    1544} & {\small  1578} &{\small 1576} &{\small 1855}   &{\small   1693 }   &{\small   1735} &{\small 1748}
 &{\small  1617}  & {\small  1516}  & {\small   1549}&{\small 1555} &{\small  1872} &{\small   1692}    &{\small 1731}&{\small 1748}  \\
{\small 1680}  & {\small   1592} &{\small   1625} &{\small 1624}  &{\small 1941}   & {\small  1760}    & {\small  1801} & {\small 1808}
&{\small  1627}  &{\small   1542}  & {\small   1574}&{\small 1575}  &{\small  1894} &{\small   1701}    &{\small 1742} &{\small 1751} \\ 
\end{tabular}
\end{center}
\end{table}
 
\begin{table}
\noindent
{\bf Table XI:} {\small Fifty eight Raman-active frequencies 
($\nu$, in ${\rm cm}^{-1}$) for ${\rm C}_{48}{\rm N}_{12}$ calculated 
by using RHF and B3LYP methods with a variety of Pople-style 
basis sets. } 
\begin{center}
\begin{tabular}{cccccccc|cccccccc}
\multicolumn{8}{c|}{\small Doubly-degenerate Modes} &
\multicolumn{8}{c}{\small Non-degenerate Modes} \\ 
\cline{2-7}\cline{10-15}
\multicolumn{4}{c}{\small B3LYP} & \multicolumn{4}{c|}{\small RHF} & 
\multicolumn{4}{c}{\small B3LYP} & \multicolumn{4}{c}{\small RHF}   \\
\cline{1-4}\cline{6-8}\cline{9-12}\cline{14-16}
{\small STO-3G} & {\small 3-21G} & {\small 6-31G} & {\small 6-31G*}  & 
{\small STO-3G} & {\small 3-21G} & {\small 6-31G} & {\small 6-31G*}  & 
{\small STO-3G} & {\small 3-21G} & {\small 6-31G} & {\small 6-31G*}  & 
{\small STO-3G} & {\small 3-21G} & {\small 6-31G} & {\small 6-31G*} \\ \hline
{\small 245 } & {\small 248} & {\small   252} &{\small 245} & {\small 263} & {\small  262 } & {\small  268} & {\small  261 } 
& {\small   264 } & {\small 264} & {\small 268} & {\small 264}  &{\small 294} & {\small  288} & {\small  291} & {\small 288 }\\ 
{\small 259 } & {\small  261} & {\small  265} &{\small 260 } & {\small   286} & {\small 281} & {\small  286} & {\small 281} 
& {\small 382 } & {\small  368} & {\small  389} &{\small 376 } & {\small 445} & {\small 406} & {\small  429} & {\small  388}\\ 
{\small 376 } & {\small  368} & {\small  387} &{\small 371 } & {\small   429} & {\small 409} & {\small  428} & {\small  413} 
& {\small  406 } & {\small 398} & {\small 415} &{\small 398 } & {\small  463} & {\small 415} & {\small  448} & {\small  425}\\ 
{\small 410 } & {\small  388} & {\small  409} &{\small 396} & {\small   470} & {\small 433} & {\small  455} & {\small  440} 
& {\small 444 } & {\small  415} & {\small 442} &{\small 424 } & {\small  510} & {\small 455} & {\small  483} & {\small 454}\\ 
{\small  447 } & {\small   429} & {\small 446} &{\small 437  } & {\small  514} & {\small 479} & {\small  497} & {\small  488} 
& {\small 471 } & {\small 458} & {\small  477}  &{\small 467  } & {\small  523} & {\small 491} & {\small  511} & {\small  489}\\
{\small 493 } & {\small  450} & {\small   482} &{\small 472 } & {\small  583} & {\small 517} & {\small  549} & {\small  540} 
& {\small 498 } & {\small  491} & {\small 500} &{\small 495 } & {\small  576} & {\small 521} & {\small  543} & {\small  524}\\  
{\small 549 } & {\small  568} & {\small  566} &{\small 551 } & {\small   604} & {\small 610} & {\small  611} & {\small  592} 
& {\small 544 } & {\small  505} & {\small 544} &{\small 510} & {\small  608} & {\small 566} & {\small  587} & {\small  575}\\ 
{\small 580 } & {\small   596} & {\small  597} &{\small 581 } & {\small  652} & {\small 647} & {\small  650} & {\small  629} 
& {\small 578 } & {\small 588} & {\small  592} &{\small 576} & {\small  646} & {\small 637} & {\small  640} & {\small 615} \\
{\small 617 } & {\small   613} & {\small  652} &{\small 627 } & {\small  725} & {\small 699} & {\small  736} & {\small  706} 
& {\small 585 } & {\small 598} & {\small  602} &{\small 588 } & {\small  650} & {\small 651} & {\small  652} & {\small 637 }\\ 
{\small 649} & {\small   629} & {\small  662} &{\small 641 } & {\small   773} & {\small 726} & {\small  759} & {\small  741} 
& {\small 621} & {\small 603} & {\small  616} &{\small 597} & {\small  723} & {\small 655} & {\small   670} & {\small  652}\\ 
{\small 665} & {\small   671} & {\small  698} &{\small 671 } & {\small  780} & {\small 762} & {\small   785} & {\small  757} 
& {\small 625} & {\small  614} & {\small 652} &{\small 627} & {\small  744} & {\small 705} & {\small   749} & {\small  716}\\ 
{\small 716} & {\small   695} & {\small  714} &{\small 699 } & {\small   831} & {\small 776} & {\small  807} & {\small  788} 
& {\small 658} & {\small  657} & {\small 684} &{\small 656} & {\small  777} & {\small  760} & {\small  784} & {\small  738}\\ 
{\small 767} & {\small   760} & {\small  768} &{\small 766 } & {\small   848} & {\small 820} & {\small  833} & {\small  821} 
& {\small  730} & {\small  690} & {\small 709} &{\small 689} & {\small  832} & {\small 779} & {\small  802} & {\small 773}\\ 
{\small 782} & {\small   776} & {\small  782} &{\small 780 } & {\small   871} & {\small 844} & {\small  853} & {\small  840} 
& {\small 768} & {\small 765} & {\small  773} &{\small 766} & {\small  867} & {\small  825} & {\small  836} & {\small  819} \\ 
{\small 841} & {\small   865} & {\small  860} &{\small 843} & {\small  935} & {\small  943} & {\small  944} & {\small  918} 
& {\small 836} & {\small 851} & {\small  844} &{\small 830} & {\small  924} & {\small  918} & {\small  919} & {\small  897}\\ 
{\small 858} & {\small   887} & {\small  877} &{\small 854} & {\small  961} & {\small  974} & {\small  966} & {\small  933} 
& {\small  856} & {\small  881} & {\small 869} &{\small 844} & {\small 953} & {\small  965} & {\small  956} & {\small  916}\\ 
{\small 1124} & {\small   1084} & {\small 1105} &{\small 1093} & {\small  1238} & {\small 1160} & {\small 1192} & {\small  1164} 
& {\small 1115} & {\small 1080} & {\small 1100} &{\small 1092} & {\small  1216} & {\small 1165} & {\small 1192} & {\small  1175}\\ 
{\small 1162} & {\small   1123} & {\small 1141} &{\small 1138 } & {\small  1281} & {\small 1197} & {\small 1225} & {\small  1215} 
& {\small 1194} & {\small 1160} & {\small 1173} &{\small 1162} & {\small  1328} & {\small 1244} & {\small 1264} & {\small 1251} \\ 
{\small 1182} & {\small   1147} & {\small 1163} &{\small 1156 } & {\small  1313} & {\small 1238} & {\small 1262} & {\small  1250} 
& {\small 1221} & {\small 1181} & {\small 1196} &{\small 1189} & {\small  1344} & {\small 1260} & {\small 1281} & {\small 1273} \\ 
{\small 1274} & {\small   1186} & {\small 1220} &{\small 1223 } & {\small  1423} & {\small 1276} & {\small 1325} & {\small  1318} 
& {\small 1316} & {\small 1241} & {\small 1267} &{\small 1265} & {\small  1400} & {\small 1335} & {\small 1374} & {\small 1348} \\    
{\small 1318} & {\small   1243} & {\small 1274} &{\small 1272 } & {\small  1476} & {\small 1341} & {\small 1383} & {\small  1377} 
& {\small 1337} & {\small 1253} & {\small 1280} &{\small 1279} & {\small  1469} & {\small 1347} & {\small 1389} & {\small  1377}\\   
{\small 1386} & {\small   1311} & {\small 1337} &{\small 1336 } & {\small  1546} & {\small 1421} & {\small 1457} & {\small  1449} 
& {\small 1367} & {\small 1293} & {\small 1322} &{\small 1320} & {\small  1498} & {\small 1375} & {\small 1424} & {\small 1380} \\    
{\small 1452} & {\small   1369} & {\small 1407} &{\small 1404 } & {\small  1608} & {\small 1468} & {\small 1517} & {\small  1503} 
& {\small 1416} & {\small 1347} & {\small 1383} &{\small 1379} & {\small  1534} & {\small 1415} & {\small 1468} & {\small  1429} \\
{\small 1466} & {\small   1388} & {\small 1420} &{\small 1420 } & {\small  1647} & {\small 1490} & {\small 1537} & {\small  1540} 
& {\small 1439} & {\small 1373} & {\small 1415} &{\small 1397} & {\small  1583} & {\small 1437} & {\small 1485} & {\small  1476}\\     
{\small 1504} & {\small   1415} & {\small 1452} &{\small 1451 } & {\small  1717} & {\small 1541} & {\small 1591} & {\small  1587} 
& {\small 1481} & {\small 1419} & {\small 1453} &{\small 1441} & {\small  1655} & {\small 1528} & {\small 1578} & {\small  1552}\\   
{\small 1571} & {\small   1489} & {\small 1517} &{\small 1515 } & {\small  1781} & {\small 1631} & {\small 1671} & {\small  1664} 
& {\small 1531} & {\small 1440} & {\small 1477} &{\small 1477} & {\small  1721} & {\small 1560} & {\small 1612} & {\small  1602}\\ 
{\small 1595} & {\small   1516} & {\small 1551} &{\small 1551 } & {\small  1831} & {\small 1670} & {\small 1710} & {\small  1720} 
& {\small 1558} & {\small 1487} & {\small 1511}  &{\small 1505}& {\small  1766} & {\small 1612} & {\small 1649} & {\small 1626} \\   
{\small 1633} & {\small   1544} & {\small 1578} &{\small 1578 } & {\small  1883} & {\small 1708} & {\small 1749} & {\small  1759} 
& {\small 1581} & {\small 1509} & {\small 1545} &{\small 1530} & {\small  1817} & {\small 1644} & {\small 1687} & {\small  1680}\\   
{\small 1662} & {\small   1578} & {\small 1610} &{\small 1610} & {\small  1916} & {\small 1735} & {\small 1778} & {\small  1788} 
& {\small 1683} & {\small 1592} & {\small 1624} &{\small 1623} & {\small  1946} & {\small 1767} & {\small 1804} & {\small 1809} \\ 
\end{tabular}
\end{center}
\end{table}

\begin{table}
\noindent
{\bf Table XII:} {\small RHF and B3LYP calculations
of  IR intensities ($I_{IR}$, in $10^{3} {\rm m}/{\rm mole}$ )
of ${\rm C}_{60}$ with the corresponding vibrational modes and
frequencies ( $\nu$, in ${\rm cm}^{-1}$).}
\begin{center}
\begin{tabular}{ccccccccccc}
Method & Mode
&\multicolumn{2}{c}{STO-3G} &\multicolumn{2}{c}{3-21G}
&\multicolumn{2}{c}{6-31G} &\multicolumn{2}{c}{6-31G*} \\
\cline{3-4}\cline{5-6}\cline{7-8}\cline{9-10}
&  & $I_{IR}$ & $\nu$      & $I_{IR}$ & $\nu$
    & $I_{IR}$ & $\nu$      & $I_{IR}$ & $\nu$ \\ \hline
B3LYP & ${\rm t}_{\rm 1u}$ &10.6  &1505 &14.0 &1454 &17.2 & 1479  &15.6& 1464   \\
&  &21.0 &1266  &9.2 & 1175  &10.1 & 1209  &8.9 &1212   \\
& &0.5  & 596  &5.9 &582  &8.2  & 587  &10.7 &570    \\
 &  &22.0 &546  &28.8 &508  &27.7 &553  &27.1 &537   \\
\\
RHF & ${\rm t}_{\rm 1u}$ &11.5  &1637  &16.5 &1553 &24.5 & 1587  &17.1 & 1549  \\
&  &25.1 &1396  &11.4 & 1245  &11.8 & 1287 & 10.6 & 1297   \\
&  &0.6  & 656  &6.0 &614  &14.7  & 623 &10.4 & 625     \\
&  &35.9 &627  &42.1 &575  &34.3 & 621  &48.0 & 599   \\
\end{tabular}
\end{center}
\end{table}
 
\begin{center}
\epsfig{file=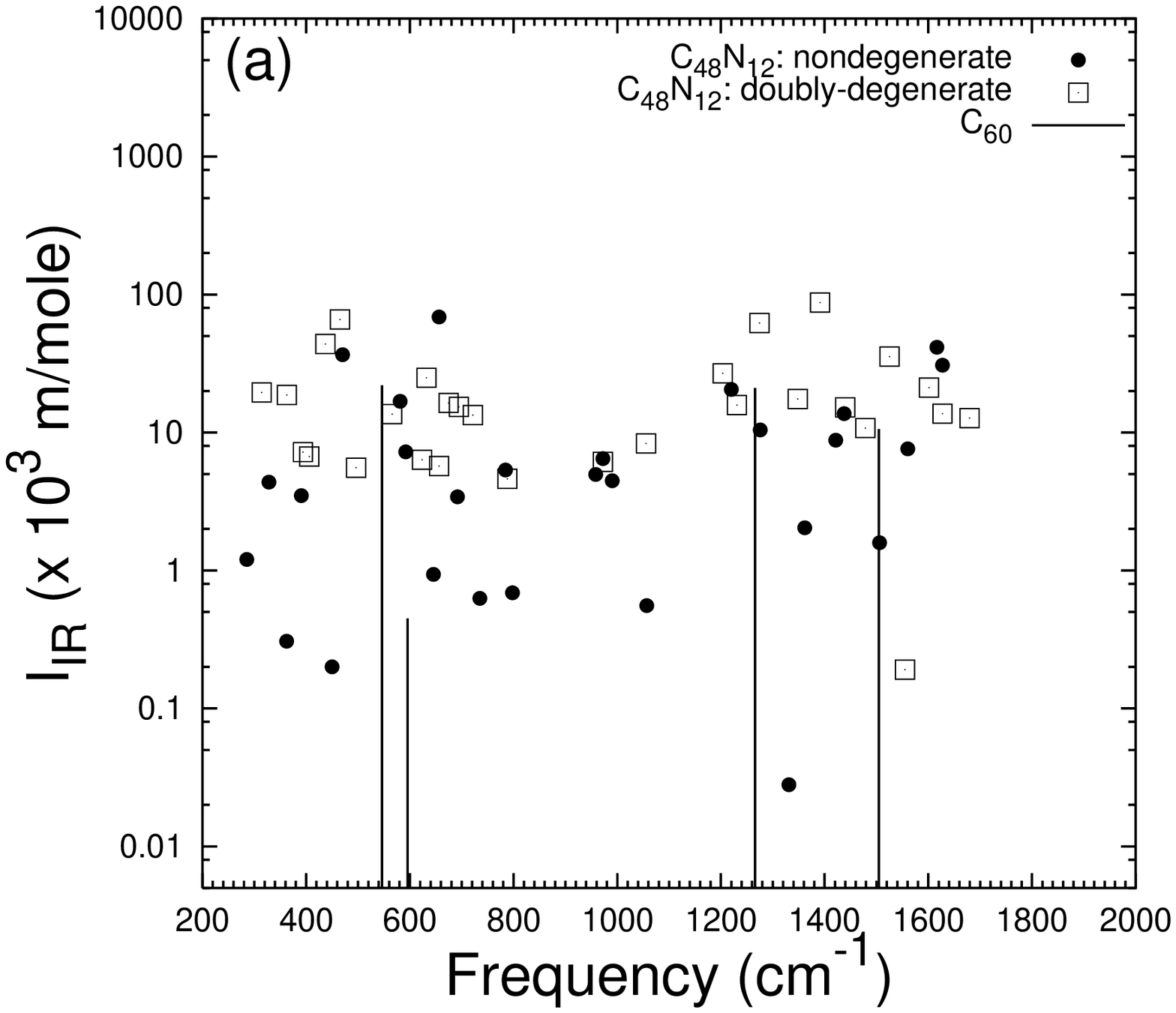,width=5cm,height=5cm}
\epsfig{file=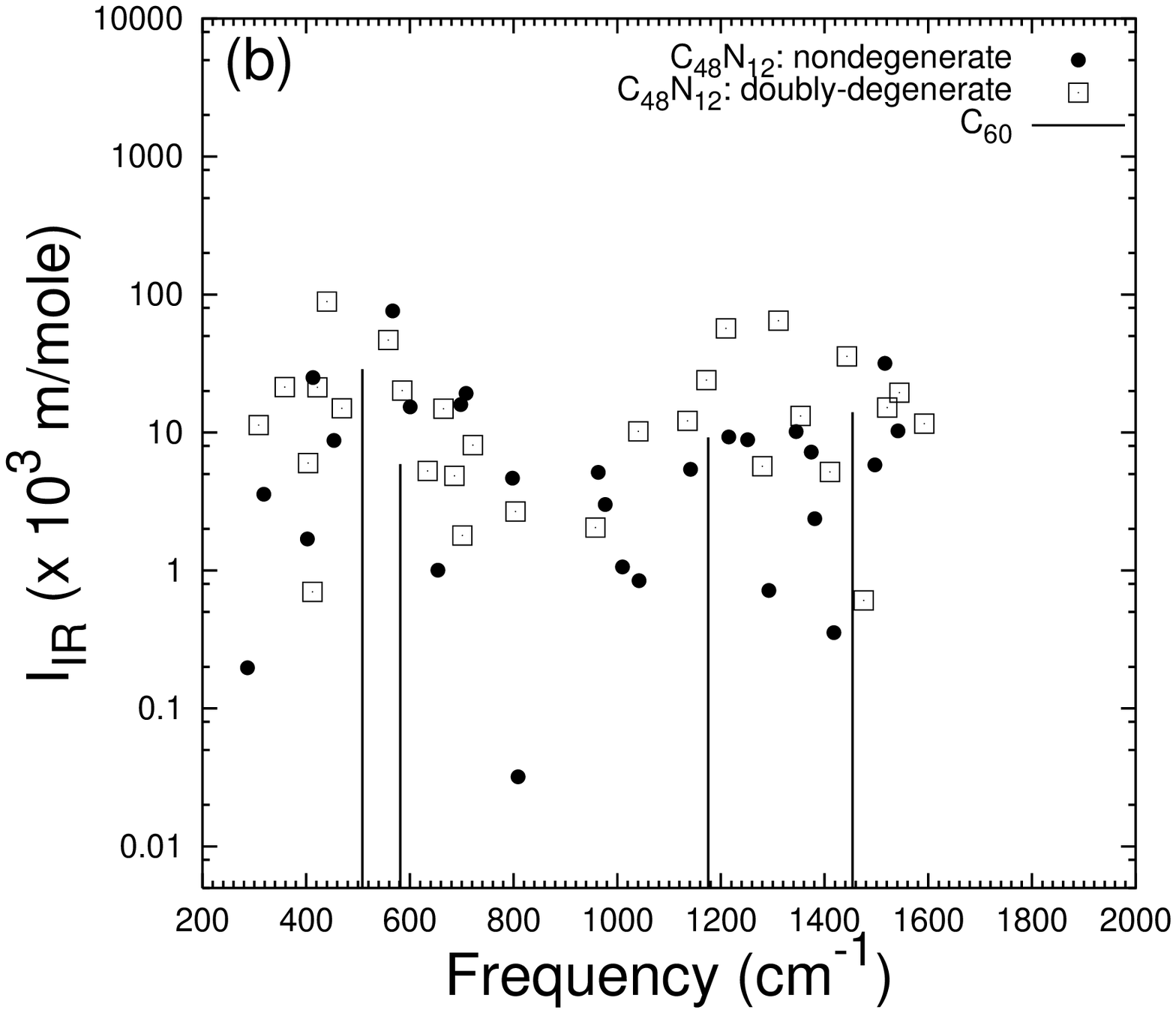,width=5cm,height=5cm}
\epsfig{file=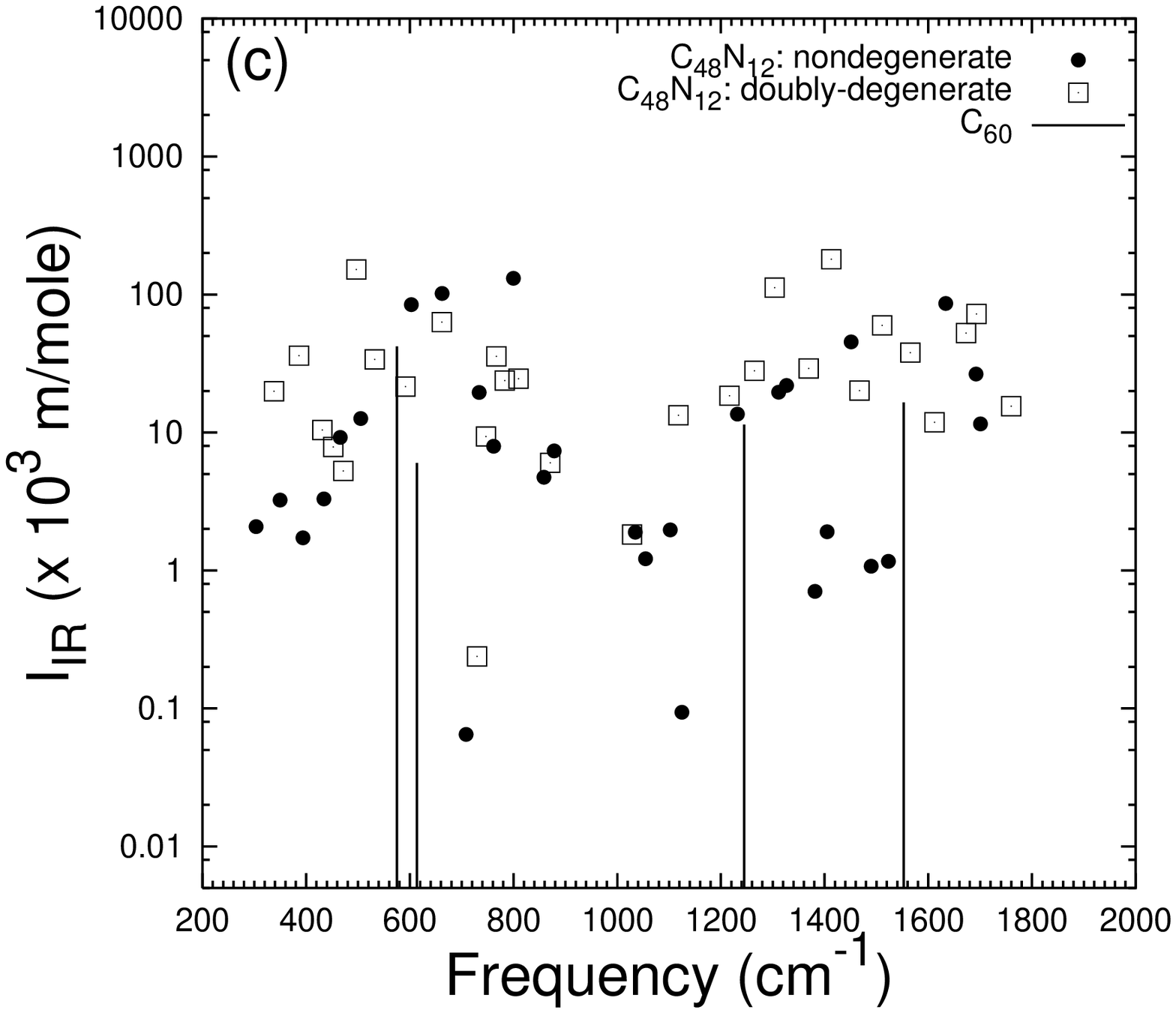,width=5cm,height=5cm}
\end{center}
 
\begin{quote}
{\bf FIG.4}:  {\sl Ab initio}  calculation of IR intensity ($I_{IR}$, 
in $10^{3}\ {\rm m/mole}$) at its corresponding frequency in 
${\rm C}_{48 }{\rm N}_{12}$: (a) B3LYP/STO-3G;
(b) B3LYP/3-21G; (c) RHF/3-21G. The solid lines are 
the calculated results for ${\rm C}_{60}$.  
\end{quote}

\begin{multicols}{2}

\noindent 
 shows collective vibration along the z-x direction, while the pentagon structure for
the high-frequency case  contracts, accompanying a large stretching of  site 3 along the
z direction.
 
\begin{center}
\epsfig{file=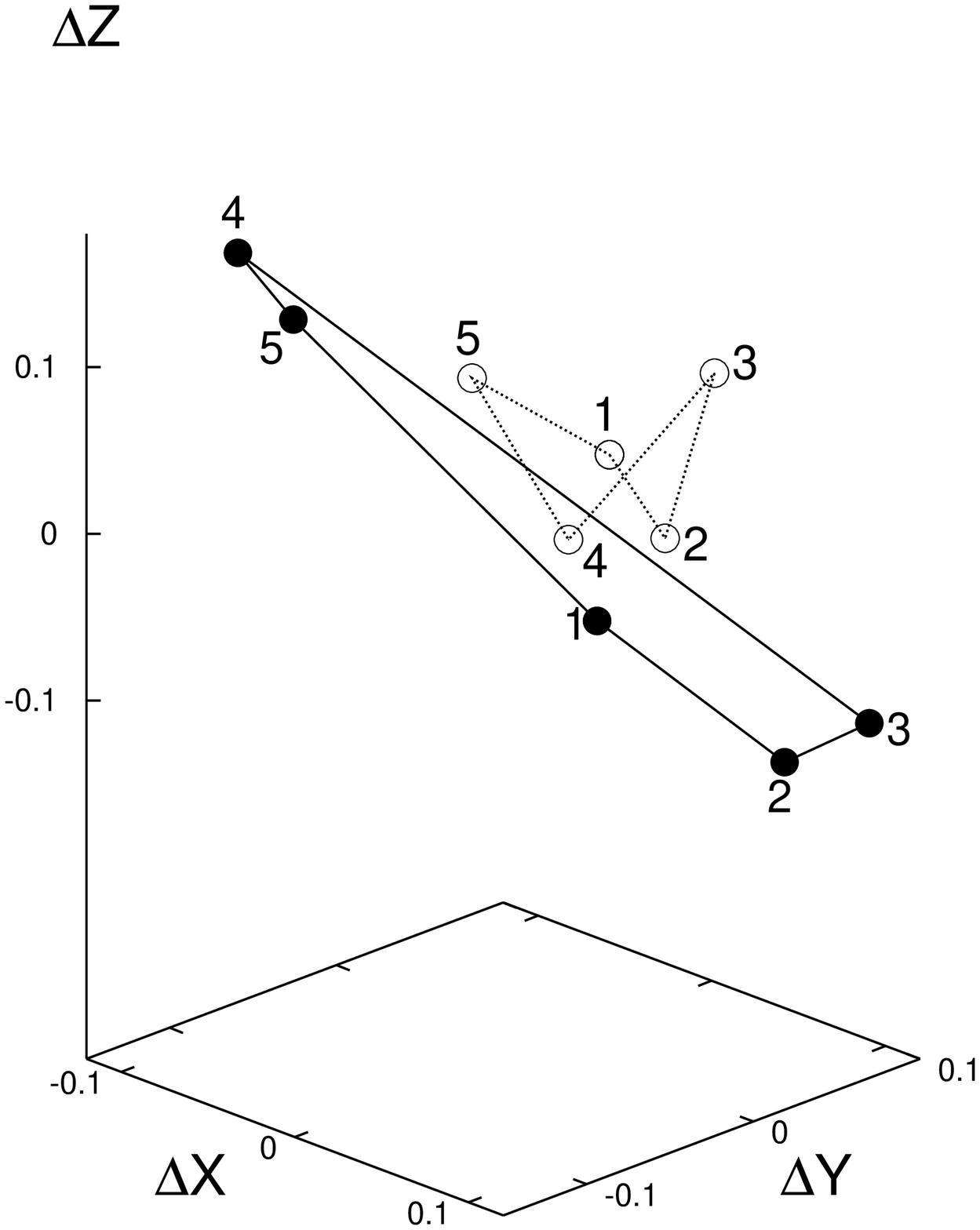,width=7cm,height=7cm}
\end{center}
\begin{quote}
 {\bf FIG.5}: The vibrational displacements of sites 1 to 5 for the strongest IR 
spectral signals of  ${\rm C}_{48}{\rm N}_{12}$ in the low-frequency (filled circles)  
and high-frequency (open circles) regions for the B3LYP/3-21G case.  
\end{quote}
 
\section{Nuclear Magnetic Shielding Tensors}
 
\subsection{Theoretical Methods}
 
There are a number of theoretical methods for calculating the
second-order magnetic response properties of molecules. It has been
shown that gauge-invariant procedures (GIPs)\cite{keith93,gauss93} are required
to predict accurately these properties within a finite basis
approximation.  In this paper, we  focus on using two GIPs,
i.e., the gauge-including atomic orbital (GIAO) procedure and the 
continuous set of gauge transformations (CSGT) procedure, to predict NMR
shielding tensors at the Hartree-Fock and DFT levels of theory.
GIAO and CSGT achieve gauge-invariance in different ways.
The GIAO method uses basis functions having explicit field dependence
\cite{london37}, whereas the CSGT method achieves the gauge-invariance
by performing a continuous set of gauge transformations. 
Ditchfield \cite{ditchfield74} first adopted the GIAO method
to perform quantum chemical NMR shift calculations and the CSGT method was
developed by Keith and Bader\cite{keith93}. In the following, we briefly
introduce the two methods. More details can be found, for example, in the works of
Cheeseman {\sl et al.}\cite{jrc96} and Wolinski {\sl et al.}\cite{giao}.

It is known that the nuclear magnetic shielding tensor can be written
as the mixed second derivative of the energy $E$ with respect to the
external magnetic field ${\rm\bf B}$ and the magnetic moment ${\rm\bf I}$
of nucleus $X$: \begin{equation}
\sigma_{ik}=\frac{\partial^{2}E}{\partial {\rm I}_{i}\partial {\rm B}_{k}},
\end{equation}
where ${\rm B}_{k}$ and ${\rm I}_{i}$ are the components of the external 
magnetic field and induced magnetic moment, respectively. The nuclear magnetic
shielding isotropy  $\sigma$ is defined as\cite{nmr1}
\begin{equation}
\sigma = (\sigma_{xx}+\sigma_{yy}+\sigma_{zz})/3,
\end{equation}
and the shielding anisotropy $\Delta\sigma$, an indication of the quality 
of the magnetic shielding tensor,  is defined  as\cite{nmr1}
\begin{equation}
\Delta\sigma = \sigma_{3} - ( \sigma_{1} + \sigma_{2} )/2,
\end{equation}
where $\sigma_{1} < \sigma_{2} < \sigma_{3} $ are the eigenvalues of the
symmetrized shielding tensor.  The nuclear magnetic shielding difference,
or say, chemical shift $\delta$, is reported in ppm (i.e., parts per million) 
and given by \cite{nmr1}
\begin{equation}
\delta=\left(\sigma^{(reference)} - \sigma^{(sample)}\right)\times 10^{6},
\end{equation}
where $\sigma^{(reference)}$ and $\sigma^{(sample)}$ denote the
shielding isotropies $\sigma$ for the reference and sample,
respectively.

In both DFT and HF theory, the expression for the nuclear magnetic
shielding tensor of nucleus X is given by \cite{jrc96}
\begin{equation}
\sigma_{ik} = <\Pi_{ik}O>+<\theta_{k}\frac{\partial O}{\partial {\rm B}_{k}}>
\end{equation}
with
\begin{eqnarray}
\Pi_{ik} &=&\langle\Phi_{\mu}\mid
\frac{ {\bf r}\bullet ({\bf r}-{\bf R}_{X})\delta_{ik}- r_{i}
({\bf r}-{\bf R}_{X})_{k}}{2c^{2}\mid {\bf r}-{\bf R}_{X}\mid^{3}}
\mid\Phi_{\nu}\rangle;\\
\theta_{k} &=& \langle\Phi_{\mu}\mid -
\frac{i[({\bf r}-{\bf R}_{X})\times\bigtriangledown]_{k}}{c\mid {\bf r}-{\bf R}_{X}\mid^{3}}
\mid\Phi_{\nu}\rangle,
\end{eqnarray}
where $c$  is the velocity of light, $O$ is the density matrix, 
$\Phi_{\mu}$ and $\Phi_{\nu}$ are 
spin orbitals, ${\bf R}_{X}$ is the position vector of nuclear X, and 
${\bf r}$ is the real space vector. In the above equation, 
$\partial O/\partial {\rm B}_{k}$ is  the derivative of the density
matrix $O$ with respect to the $k$th component of the magnetic field 
${\rm\bf B}$ and  is obtained via solution of the coupled-perturbed 
equations\cite{jrc96} for the appropriate perturbation. As gauge-invariance 
is achieved in different ways, the GIAO and CSGT methods differ at this 
point in the formation of the coupled-perturbed equations\cite{jrc96}.

\end{multicols}

\begin{table}
\noindent
{\bf Table XIII:}  {\small RHF and B3LYP calculations of the absolute
isotropy ($\sigma$, in ppm) and anisotropy ($\Delta\sigma$, in ppm)  of
the carbon and nitrogen shielding tensors with a variety of Pople-style basis sets
 for ${\rm C}_{48}{\rm N}_{12}$ aza-fullerene,  ${\rm C}_{60}$ and tetramethylsilane (TMS) by using the GIAO method. 
Numbers in the parenthesis for ${\rm C}_{60}$  are the relative errors of the calculated 
$^{13}{\rm C}$ NMR shift $\delta$ to the NMR chemical shift $\delta^{exp.}=142.7\ {\rm ppm}$ measured by Taylor
{\sl et al.}\cite{taylor90}.}

\begin{center}
\begin{tabular}{cccccccccccc}
  & &    &    & \multicolumn{2}{c}{STO-3G}&
\multicolumn{2}{c}{3-21G} &\multicolumn{2}{c}{6-31G} &
\multicolumn{2}{c}{6-31G*} \\
\cline{5-6}\cline{7-8}\cline{9-10}\cline{11-12}
Method & Molecule & Site Numbers $\{ n_{i} \}$ & Nuclei & $\sigma$ & $\Delta\sigma$ &$\sigma$ & $\Delta\sigma$  &$\sigma$ & $\Delta\sigma$
& $\sigma$ & $\Delta\sigma$  \\ \hline
 RHF&${\rm C}_{48}{\rm N}_{12}$ & \{1, 13, 16, 31, 38, 51\}  & $^{13}{\rm C}$    &91.5   &155.5 &65.9 &155.6 & 44.0  & 173.0  &48.5   & 168.3   \\
 & & \{2, 12, 29, 32, 37, 52\}  & $^{13}{\rm C}$    &109.8  &156.7   &89.8  &149.2  & 67.9  & 169.2  &67.7   & 168.2  \\
 & &\{3, 11, 28, 33, 36, 53\}  & $^{13}{\rm C}$    &116.8  &132.2    &99.2  &123.5  & 76.6  & 141.2  &88.5   & 126.9 \\ 
 & & \{4, 15, 27, 34, 40, 54\}  & $^{13}{\rm C}$    &103.5  &141.9  &75.8  & 139.0 & 57.9  & 154.9  &46.9   & 156.7 \\
 & &\{ 5, 14, 30, 35, 39, 55\} & $^{15}{\rm N}$     &193.7  &134.8    &142.7  & 143.8 & 114.4 & 171.5  &140.7   & 128.5  \\  
 & &\{6, 18, 24, 42, 48, 58\}  & $^{13}{\rm C}$    &120.2  &104.8    &101.6  &89.8  & 78.2  & 110.5  &89.6   &  82.5  \\ 
 & & \{7, 19, 23, 43, 47, 57\}  & $^{13}{\rm C}$     &106.2  &135.9   &80.6  &131.0  & 59.4  & 152.2  &54.2   & 148.7\\
& &\{8, 20, 22, 44, 46, 56\}  & $^{13}{\rm C}$    &111.5  &140.9    &94.6  &126.1  & 70.7  & 150.8  &82.9   & 131.1 \\
 & &\{9, 21, 26, 45, 50, 60\}  & $^{15}{\rm N}$    &182.7  &151.6    &132.6  &167.7  & 95.3  & 202.6  &121.7   & 170.4 \\   
  & & \{10, 17, 25, 41, 49, 59\} & $^{13}{\rm C}$    &104.9  &125.2   &80.0  &117.0  & 59.5  & 133.5  &55.1   & 126.9 \\
  &${\rm C}_{60}$ & \{1, 2, 3, 4, 5, 6, ..., 60\} & $^{13}{\rm C}$ &101.8  &156.6  &74.0  &162.7  & 54.1  & 180.1  & 54.7  & 178.8   \\
  &TMS &  carbon site   & $^{13}{\rm C}$ &238.5  & 5.6  &208.3  & 16.9 & 200.8 & 21.4   & 195.1  &17.5   \\
  &${\rm NH}_{3}$ & nitrogen site & $^{15}{\rm N}$  &306.9  & 9.2 &271.1   &17.5  &264.4  &18.2  &260.8  &17.4   \\
 &${\rm C}_{60}$ &\multicolumn{2}{c}{Calculated $^{13}{\rm C}$ NMR shift $\delta$}  &\multicolumn{2}{c}{ 136.7 (4.2\%)}
                                                           &\multicolumn{2}{c}{ 134.3 (5.9\%)}  &\multicolumn{2}{c}{ 146.7 (2.8\%)}
                                                            &\multicolumn{2}{c}{ 140.4 (1.6\%)} \\
  \\
 B3LYP&${\rm C}_{48}{\rm N}_{12}$ &\{1, 13, 16, 31, 38, 51\}  & $^{13}{\rm C}$    & 103.9 &109.5 &70.9 & 117.5   &53.3 & 128.0  & 51.9  & 130.4  \\
 & &\{2, 12, 29, 32, 37, 52\}  & $^{13}{\rm C}$    & 114.6  &121.9 & 84.6 & 123.6   &66.4  &138.1 &62.8   & 138.9   \\
 & &\{3, 11, 28, 33, 36, 53\}  & $^{13}{\rm C}$    & 120.6  &101.1 & 91.8 & 98.6  &74.5 &110.9  &72.9   & 107.3 \\
& &\{4, 15, 27, 34, 40, 54\}  & $^{13}{\rm C}$    & 113.3  &103.6 &82.8 & 104.6 &66.3 & 117.1  &61.8   & 113.9  \\
& &\{ 5, 14, 30, 35, 39, 55\} & $^{15}{\rm N}$    & 173.2 &126.2  & 116.4 & 143.2   &92.8  &155.4 &103.8  &  139.3 \\
 & &\{6, 18, 24, 42, 48, 58\}  & $^{13}{\rm C}$    & 123.5  &70.4  & 92.5 & 66.5   &76.1  &76.4  & 74.3  & 63.0   \\
 & &\{7, 19, 23, 43, 47, 57\}  & $^{13}{\rm C}$    & 114.1  &96.4  & 80.6 & 100.2  &63.0  &76.4  & 57.4  & 114.0 \\
 & &\{8, 20, 22, 44, 46, 56\}  & $^{13}{\rm C}$    & 119.5  &101.7 & 90.5 & 104.7  &73.8  &116.2 &72.3   & 116.2 \\ 
 & &\{9, 21, 26, 45, 50, 60\}  & $^{15}{\rm N}$    & 171.7  &121.9 & 110.4 & 142.2    &84.0 &157.5 &91.2   & 146.7 \\  
 & &\{10, 17, 25, 41, 49, 59\} & $^{13}{\rm C}$    & 113.9  &121.9 & 82.0 & 84.6   & 65.6 &94.4  &61.7   & 90.0 \\
 &${\rm C}_{60}$ &\{1, 2, 3, 4, 5, 6, ..., 60\} & $^{13}{\rm C}$ & 106.1 &125.8 & 70.2 & 139.3  &53.6 &152.2 &50.5   & 154.2  \\
 &TMS &  carbon site   & $^{13}{\rm C}$ &225.3 & 9.6 &197.6   &20.0   & 188.3  & 25.2  &183.8   &22.3   \\
 &${\rm NH}_{3}$ & nitrogen site &  $^{15}{\rm N}$ &291.3 & 7.3 &262.6  &18.9 &256.8  &21.0 &254.9  &19.3  \\
 &${\rm C}_{60}$ &\multicolumn{2}{c}{Calculated $^{13}{\rm C}$ NMR shift $\delta$}  &\multicolumn{2}{c}{ 119.2 (16.5\%)}
                                         &\multicolumn{2}{c}{127.4  (10.7\%)  }  &\multicolumn{2}{c}{ 134.7 (5.6\%)}
                                         &\multicolumn{2}{c}{ 133.3 (6.6\%)} \\
\\
Experiment & ${\rm C}_{60}$  & \multicolumn{8}{c}{$^{13}{\rm C}$ NMR chemical shift $\delta$ from Taylor {\sl et al.}
\cite{taylor90}:}&\multicolumn{2}{c}{142.7}\\
  & TMS &\multicolumn{8}{c}{absolute shielding isotropy  $\sigma$ for $^{13}{\rm C}$ from Jameson {\sl et al.}
\cite{jameson87}:}&\multicolumn{2}{c}{188.1}\\
 & ${\rm NH}_{3}$ &\multicolumn{8}{c}{absolute shielding isotropy ($\sigma$)
 /anisotropy ($\Delta\sigma$) for $^{15}{\rm N}$ from Ref.\cite{sgk75,jameson81}:}&\multicolumn{2}{c}{264.5/20}\\
\end{tabular}
\end{center}
\end{table}

\begin{multicols}{2}
 
\subsubsection{GIAO Method}
 
The GIAO method uses the following explicit field-dependent basis functions for
calculating the magnetic shielding tensor:
\begin{equation}
\mid\Phi_{\mu}({\rm\bf B})\rangle = e^{-i({\rm\bf B}
\times {\rm\bf r}_{\mu})\bullet {\bf r}/(2c)}
\mid\Phi_{\mu}(0)\rangle,
\end{equation}
where ${\rm\bf r}_{\mu}$ is the position vector of basis function
$\Phi_{\mu}$ and $\Phi_{\mu}(0)$ denotes the usual field-independent
basis functions. In this method, three sets of the coupled-perturbed
equations\cite{jrc96} are solved, one for each direction of
the magnetic field.

\subsubsection{CSGT Method}
 
In the CSGT method, the gauge-invariance is achieved from accurate
calculations of the induced first-order electronic current density
${\bf J}^{(1)}({\bf r})$ by performing a gauge transformation for
each point in space. The nuclear magnetic shielding tensor is expressed
in terms of ${\bf J}^{(1)}({\bf r})$, i.e.,
\begin{equation}
\sigma_{ik}= -\frac{1}{{\rm B}c}\int\left[\frac{{\bf R}_{X}
\times {\bf J}_{k}^{(1)}({\bf r})}{R_{X}^{3}}\right]_{i} d{\bf R}_{X},
\end{equation}
 
\end{multicols}
 
\begin{table}

\noindent 
{\bf Table XIV:} {\small RHF and B3LYP calculations of the absolute 
isotropy ($\sigma$, in ppm) and anisotropy ($\Delta\sigma$, in ippm)  of 
the carbon and nitrogen shielding  tensors with a variety of Pople-style basis sets 
 for ${\rm C}_{48}{\rm N}_{12}$ aza-fullerene,  
${\rm C}_{60}$ and tetramethylsilane (TMS) by using CSGT methods. Numbers 
in the parenthesis for ${\rm C}_{60}$  are the relative errors of the calculated
$^{13}{\rm C}$ NMR shift $\delta$  to the NMR chemical shift $\delta^{exp.}=142.7\ {\rm ppm}$ 
measured by Taylor {\sl et al.}\cite{taylor90}.} 
 
\begin{center}
\begin{tabular}{cccccccccccc}
  & &    &    & \multicolumn{2}{c}{STO-3G}&\multicolumn{2}{c}{3-21G} &\multicolumn{2}{c}{6-31G} & \multicolumn{2}{c}{6-31G*} \\
\cline{5-6}\cline{7-8}\cline{9-10}\cline{11-12}
 Method & Molecule & Site Numbers $\{ n_{i} \}$ & Nuclei & $\sigma$ & $\Delta\sigma$ &$\sigma$ & $\Delta\sigma$  &$\sigma$ & $\Delta\sigma$
& $\sigma$ & $\Delta\sigma$  \\ \hline
RHF & ${\rm C}_{48}{\rm N}_{12}$ & \{1, 13, 16, 31, 38, 51\}  &  $^{13}{\rm C}$   &45.3   &125.0  &60.7 & 151.5 &30.7  &168.9  & 41.3  & 170.1   \\
 & &\{2, 12, 29, 32, 37, 52\}  &  $^{13}{\rm C}$     &55.8   &127.0 &84.1 &143.8   &54.0  &162.1  & 60.4  & 168.3  \\ 
& &\{3, 11, 28, 33, 36, 53\}  &  $^{13}{\rm C}$     &60.0   &114.0 &91.2 &122.4   &60.9  &139.1  & 78.9  & 129.8 \\ 
 & &\{4, 15, 27, 34, 40, 54\}  &  $^{13}{\rm C}$     &52.9   &119.5 &68.8   &138.7 &41.1  &155.5  & 39.2  & 160.4  \\
 & &\{ 5, 14, 30, 35, 39, 55\} &  $^{13}{\rm N}$     &78.0  &134.0  &120.4 &154.3  &86.3  &176.5    & 128.2  & 136.4  \\ 
& &\{6, 18, 24, 42, 48, 58\}  &   $^{13}{\rm C}$     &63.8   &97.9  &89.8 &96.5    &59.3  &115.5   & 79.8  & 87.3   \\ 
 & &\{7, 19, 23, 43, 47, 57\}  &  $^{13}{\rm C}$     &55.5   &113.5 &72.2 &130.9   &43.2  &149.5  & 46.5  & 150.5 \\
 & &\{8, 20, 22, 44, 46, 56\}  &  $^{13}{\rm C}$     &58.6   &118.0 &84.9 &125.5   &52.4  &150.1  & 73.7  & 135.2 \\ 
 & &\{9, 21, 26, 45, 50, 60\}  &  $^{13}{\rm N}$    &74.0   &144.8 &116.0 &170.8  &72.2  &203.6  & 112.2  & 173.1 \\
 & &\{10, 17, 25, 41, 49, 59\} &   $^{13}{\rm C}$     &55.3   &107.8 &73.1 &118.7   &43.3  &136.9  & 47.3  & 129.5  \\
 &${\rm C}_{60}$ &\{1, 2, 3, 4, 5, 6, ..., 60\} &  $^{13}{\rm C}$  & 50.2 &128.8 &71.2 &156.1 &43.3  &172.7    & 48.9   &178.6   \\
 &TMS &  carbon site   &  $^{13}{\rm C}$  & 131.9 & 9.5 &191.0  &6.3 &185.5  &13.1    & 190.6 & 13.7 \\
  &${\rm NH}_{3}$ & nitrogen site &  $^{13}{\rm N}$  &148.1  &18.6  &200.2   &5.1  &209.2  & 6.3 &228.4  & 6.6  \\   
  &${\rm C}_{60}$ &\multicolumn{2}{c}{Calculated $^{13}{\rm C}$ NMR shift $\delta$}  &\multicolumn{2}{c}{ 81.7 (42.7\%)}
     &\multicolumn{2}{c}{119.8 (16.0\%)} &\multicolumn{2}{c}{ 142.2 (0.4\%)}&\multicolumn{2}{c}{ 141.7 (0.7\%)} \\ 
 \\
 B3LYP&${\rm C}_{48}{\rm N}_{12}$ &\{1, 13, 16, 31, 38, 51\}  &  $^{13}{\rm C}$     &58.6  &86.4 & 66.0 & 114.1    &39.1  &126.6   & 45.7  & 131.8  \\
& &\{2, 12, 29, 32, 37, 52\}  &  $^{13}{\rm C}$     &66.1   &94.8 & 81.7 & 119.3  &53.9   &133.5   &57.2   &139.6   \\ 
 & &\{3, 11, 28, 33, 36, 53\}  &  $^{13}{\rm C}$     &69.3   &83.5 &87.2 & 99.0  &60.3   &111.1  &66.3   &109.4  \\ 
& &\{4, 15, 27, 34, 40, 54\}  &  $^{13}{\rm C}$     &64.6  &84.8  & 75.8 & 104.0 &49.1   &118.5  & 54.7  & 117.6 \\
 & &\{ 5, 14, 30, 35, 39, 55\} &  $^{13}{\rm N}$     &78.5  &116.6 & 104.9 & 142.0 &74.6  &152.7   &96.8  &142.4  \\
 & &\{6, 18, 24, 42, 48, 58\}  &  $^{13}{\rm C}$     &71.9   &63.6 &84.1 & 72.0 &58.7   &81.9    &66.9   &66.5    \\ 
 & &\{7, 19, 23, 43, 47, 57\}  &  $^{13}{\rm C}$     &66.5   &77.7 & 73.8 & 99.7  &47.5   &112.2   &51.0   &115.3  \\
 & &\{8, 20, 22, 44, 46, 56\}  &  $^{13}{\rm C}$     &70.4   &81.9 & 83.5 & 101.7  &56.7   &115.8   &65.7   &118.7  \\
 & &\{9, 21, 26, 45, 50, 60\}  &  $^{13}{\rm N}$     &81.6   &112.8 &103.4 &140.2  & 69.3 &156.6 &86.0   &148.5  \\ 
 & &\{10, 17, 25, 41, 49, 59\} &  $^{13}{\rm C}$     &66.9   &71.6 & 76.1 & 88.1  &49.5   &100.1   &54.8   &93.8  \\
 &${\rm C}_{60}$ &\{1, 2, 3, 4, 5, 6, ..., 60\} &  $^{13}{\rm C}$  &58.2  &100.3 &69.2 &134.7 &43.1   &148.4   & 45.9   & 149.2  \\
  &TMS &  carbon site   &  $^{13}{\rm C}$  & 124.9 & 10.0  &181.7 &9.0   &175.0   &16.5 &181.6  &18.0   \\
  &${\rm NH}_{3}$ & nitrogen site &  $^{13}{\rm N}$  &137.7  &27.1  &190.0   &9.6  &200.4  &6.0  &221.5  &6.5   \\
  &${\rm C}_{60}$ &\multicolumn{2}{c}{Calculated $^{13}{\rm C}$ NMR shift $\delta$}  &\multicolumn{2}{c}{ 66.7 (53.3\%)}
   &\multicolumn{2}{c}{ 112.5 (21.2\%)} &\multicolumn{2}{c}{ 131.9 (7.6\%)}&\multicolumn{2}{c}{ 135.7  (4.9\%)} 
\end{tabular}
\end{center}
\end{table}

\begin{multicols}{2}

\noindent 
where $B$ is the magnitude of magnetic field. Since the basis
functions do not depend on the
magnetic field, only six sets of the coupled-perturbed equations\cite{jrc96} are
left and solved (three for the
components of the angular momentum perturbation using any single gauge
origin, and three for the linear momentum perturbation resulting from
any single shift in gauge origin). Furthermore, the shielding tensors are
obtained via Becke's multi-center numerical integration scheme\cite{becke22}

\subsection{Results}
 
The above mentioned GIAO and CSGT methods have been implemented into 
the Gaussian 98 program\cite{nist,gaussian} and all  calculations in this section 
are performed using this program.
 
Here we perform RHF and B3LYP hybrid DFT calculations of the nuclear magnetic
shielding tensor  of carbon and nitrogen atoms in
${\rm C}_{48}{\rm N}_{12}$, ${\rm C}_{60}$, and
tetramethylsilane (TMS\cite{jameson87}) by using GIAO and CSGT methods 
with STO-3G, 3-21G, 6-31G and 6-31G* basis sets.
Calculation performed with a specific basis set and {\sl ab initio} method use
geometry optimized with the same basis set and {\sl ab initio} method (see section II). 
Detailed results are summarized in Table XIII and XIV. Since no present functional 
includes a magnetic field dependence, the DFT methods do not provide 
systematically better shielding results than RHF\cite{gaussian}. Nevertheless, 
we also list the DFT results in Table XIII and XIV for comparison.  The 
isotropic NMR chemical shifts $\delta$ relative to that of TMS are also shown
in Table XIII and XIV.  It is seen that the NMR shielding tensors for
${\rm C}_{48}{\rm N}_{12}$ are separated into eight groups for carbon atom and
two groups for nitrogen atoms, i.e., eight $^{13}$C and two $^{15}$N NMR 
spectral signals are predicted. In contrast, ${\rm C}_{60}$ has only one 
$^{13}$C NMR signal which is in agreement with experiment\cite{taylor90}. 
The experimental values of the absolute shielding constant $\sigma$ for 
$^{15}$N in ${\rm NH}_{3}$ and $^{13}$C in TMS, as shown in Table XIII and 
XIV,  are   $264.5\ {\rm ppm}$\cite{sgk75,jameson81} and $188.1\ {\rm ppm}$
\cite{jameson87}, and our best calculated results for those reference 
materials are in good agreement with the experiments.

Table XIII and XIV also demonstrate the convergence of the GIAO and CSGT 
methods with respect to basis set for absolute shielding constants calculated 
with RHF and B3LYP hybrid DFT methods. For each first principles 
method, the shielding constants calculated with GIAO and CSGT methods are 
found to converge to almost the same values for the large basis set 6-31G*. 
NMR chemical shifts, as opposed to the absolute shielding constants $\sigma$,  
are measured with high accuracy in applications of NMR spectroscopy. On 
the other hand, calculated chemical shifts are in better agreement with 
experiment as relative differences are better represented\cite{jrc96}.  Hence, 
given the absolute shielding constants for $^{13}$C and $^{15}$N in reference 
materials, i.e., TMS and ${\rm NH}_{3}$ shown in Table XIII and XIV,  we are 
able to calculate the chemical shifts $\delta$ and compare them with the 
experiments. For ${\rm C}_{60}$, we find that the NMR chemical shifts $\delta$ calculated
with the CSGT method at the RHF/6-31G and RHF/6-31G* levels are in agreement 
with  experiment ($\delta^{exp} = 142.7 \ {\rm ppm}$ measured by 
Taylor {\sl et al.}\cite{taylor90}). This suggests that CSGT method would 
predict NMR spectral signals much better than GIAO.
However, CSGT costs  more CPU times than GIAO on the same machine, 
for example, about 5 hours more for RHF/6-31G* and 2 hours for  B3LYP/6-31G* 
in the CSGT method.

To predict accurate NMR chemical shifts for large molecules, 
it is necessary to assess the accuracy of the available {\sl ab initio} 
methods by employing lower levels of theory and  by using basis sets as 
small as possible\cite{jrc96}.  The results in Table XIII and XIV  indicate that the RHF 
method yields better chemical shifts than the B3LYP hybrid DFT method, 
especially, for GIAO method and the minimum basis set STO-3G. As shown in 
Table XIII,  the $^{13}$C chemical shift in ${\rm C}_{60}$ differs from 
experiment by 23.5 ppm at the B3LYP level, while it differs by only 6 ppm 
at the RHF level of theory. Also it is noted that RHF makes positive and negative 
errors, while B3LYP  makes only positive errors.
 
In addition, nuclear magnetic shielding anisotropies $\Delta\sigma$ are 
reported in Table XIII and XIV. Although these properties usually cannot be 
determined 
experimentally in the gas phase (except in cases where the high symmetry of 
the molecule together with the calculated diamagnetic contribution to the 
shielding tensor allows the determination of the anisotropy from the spin 
rotation\cite{gauss95}), these values are of interest. Anisotropies 
can be determined in solid state NMR experiments and calculations are often 
important for a correct assignment. From Table XIII and XIV, we find that 
the  shielding anisotropies $\Delta\sigma$ for $^{15}$N in 
${\rm NH}_{3}$ calculated with GIAO method with both {\sl ab initio} theories are in 
good agreement with experiment, while those calculated with CSGT agree poorly 
with experiment.

\section{summary and outlook}

In summary, we have performed {\sl ab initio} calculations of the structures, 
electronic properties, vibrational frequencies, IR intensities, NMR shielding 
tensors, linear polarizabilities, first- and second-order hyperpolarizabilities of the 
${\rm C}_{48}{\rm N}_{12}$ aza-fullerene. Calculated results of ${\rm C}_{48}{\rm N}_{12}$ 
are compared with those of ${\rm C}_{60}$ at the same level of theory. It is 
found that this aza-fullerene has some remarkable 
features, which are different from and can compete with ${\rm C}_{60}$. 
The detailed studies of ${\rm C}_{60}$ show the importance of electron 
correlations and the choice of basis sets
in the {\sl ab initio} calculations. Our best  results for
${\rm C}_{60}$ obtained with the B3LYP hybrid DFT method are in excellent 
agreement with experiment and demonstrate the needed efficiency 
and accuracy of this method for obtaining quantitative
information on the structural, electronic and vibrational properties of
these materials.

Laser sources are widely used in the laboratory 
and industry. However, they are a potential hazard for eyes, 
thermal cameras and other optical sensors\cite{book4b,book4c,vivien02a}. 
Development of  optical limiters which can 
suppress undesired radiation and effectively decrease transmittance at 
high intensity or fluence is necessary. An ideal optical limiter
\cite{book4b,vivien02a} should have reasonable linear transmittance at 
low input fluence (at least of 70\%), protecting  optical sensors or 
eyes against laser pulses of any  wavelength and pulse duration, and 
its output energy must remain  below the optical damage 
threshold  of sensors or eyes at high fluences.  ${\rm C}_{60}$ and 
fullerene derivatives are good candidates for optical limiting 
applications\cite{book4b,tutt91a,opl00}. In this paper, we found that the average 
second-order hyperpolarizability of ${\rm C}_{48}{\rm N}_{12}$ is about 
55\% larger than that of ${\rm C}_{60}$. Hence, it is expected that 
${\rm C}_{48}{\rm N}_{12}$ is also a good candidate for optical limiting 
applications.   

Non-linear optical (NLO) materials have vast technological applications 
in telecommunications, optical data storage and optical information 
processing. The search for materials with  such properties is the 
subject of intense research\cite{book4,book4b,nlo00}. Organic molecules, 
which have  high NLO response, could 
be designed by linking appropriate electron-donor and acceptor groups 
with a spacer made of one or several units of conjugated 
molecules such as a polyene or an aromatic chain\cite{nlo00}.  Based on such 
donor-acceptor model, we can link donor ${\rm C}_{48}{\rm N}_{12}$ and acceptor  
${\rm C}_{60}$ with a polyene or an aromatic chain and design one kind of organic 
molecules. In such organic molecules, charge would migrate from 
${\rm C}_{48}{\rm N}_{12}$ to ${\rm C}_{60}$ upon electronic excitation,
 giving rise to a large dipole moment along the direction connecting the donor/acceptor pair. Thus, 
large NLO response is expected in these donor/acceptor-based molecules.

Carbon nanotubes are promising materials for building electronic devices,
in particular, field effect transistors (FETs) \cite{book4c,rmartel98a}. However, 
single-walled  carbon nanotube (SWNT) FETs built from as-grown tubes are unipolar 
$p$-type, i.e., there is no electron current flow even at large positive 
gate biases. This behavior suggests the presence of a Schottky barrier at 
the metal-nanotube contact. Obviously, the
capability to produce $n$-type transistors is important technologically, as
it allows the fabrication of nanotube-based complementary logic devices
and circuits \cite{circuit7,circuit8}.  Very recently, Bockrath {\sl et al.} \cite{bockrath00a}
have reported a controlled chemical doping of individual semiconducting SWNT 
ropes  by reversibly intercalating and deintercalating potassium. 
Potassium doping could change the carriers in the ropes from holes to
electrons\cite{bockrath00a}.  Their experiments open the way toward other experiments that 
require controlled doping such as making nanoscale $p$-$n$ junctions. 
As shown in this paper, ${\rm C}_{48}{\rm N}_{12}$ is a good electron 
donor. We found that incorporating ${\rm C}_{48}{\rm N}_{12}$
into a (10,10) semiconducting SWNT\cite{book1} would result in  
-0.26 e charge\cite{xierh02b} on the nanotube forming a $n$-type 
SWNT-based transistor. 

To obtain molecular rectification, the LUMO of the acceptor
should lie at or slightly above the Fermi level of the electrode and above
the HOMO of the donor\cite{review1}. Hence it is important to search for 
desired donor/acceptor pairs which satisfy those conditions. The 
acceptor/donor pair, ${\rm C}_{48}{\rm B}_{12}$/${\rm C}_{48}{\rm N}_{12}$,
actually is demonstrated to be an ideal candidate for molecular electronics\cite{xierh02b}. 
For example, we have shown that a rectifier formed by ${\rm C}_{48}{\rm B}_{12}$/${\rm C}_{48}{\rm N}_{12}$ 
pair shows a perfect rectification behavior \cite{xierh02b}. 

As shown before, ${\rm C}_{48}{\rm B}_{12}$ and ${\rm C}_{48}{\rm N}_{12}$ fcc solids 
are semiconducting materials.  Because both ${\rm C}_{48}{\rm B}_{12}$ and ${\rm C}_{48}{\rm N}_{12}$ 
molecules have opposite electronic  polarizations, it is possible to use them 
to build semiconducting materials with opposite electronic polarizations.

Therefore, ${\rm C}_{48}{\rm N}_{12}$ could have potential applications as semiconductor 
components, nonlinear optical materials,  and possible building materials for molecular 
electronics and  photonic devices. Efficiently synthesizing ${\rm C}_{48}{\rm N}_{12}$ 
would be of great experimental interest within reach of today's technology.

\section*{acknowledgements}

We thank Dr. Denis A. Lehane and Dr. Hartmut Schmider for their 
technical help. One of us (R. H. X.) thanks the HPCVL (High 
Performance Computing Virtual Laboratory) at Queen's University 
for the use of its parallel supercomputing facilities. L.J. gratefully acknowledges 
the Danish Research Training Council for financial support. J.Z acknowledges  support 
from the University Research Council of University of North Carolina at Chapel Hill. 
VHS gratefully acknowledges support from the Natural Science and Engineering 
Research Council of Canada(NSERC).

\end{multicols}

\begin{references}
\bibitem{nist} Use of this software does not constitute an endorsement
or certification by NIST.
\end{references}
\end{document}